\documentclass[a4paper,fleqn,usenatbib]{mnras}

\usepackage[T1]{fontenc}
\usepackage{ae,aecompl}

\usepackage{graphicx}	
\usepackage{amsmath}	
\usepackage{amssymb}	

\newcommand{\Sersic}{S\'{e}rsic}

\defcitealias{2011ApJS..196...11S}{S11}
\defcitealias{2015MNRAS.447.2753T}{T15}
\defcitealias{2015MNRAS.454.1886S}{S15}

\title[Illustris galaxies in the SDSS - II]{Galaxies in the Illustris simulation as seen by the Sloan Digital Sky Survey - II: Size-luminosity relations and the deficit of bulge-dominated galaxies in Illustris at low mass}

\author[Bottrell et al.]{
Connor Bottrell,$^{1}$\thanks{E-mail: connor.bottrell@gmail.com}
Paul Torrey,$^{2}$\thanks{Hubble Fellow}
Luc Simard,$^{3}$
and Sara L. Ellison$^{1}$
\\
\scriptsize
$^{1}$University of Victoria, Department of Physics and Astronomy, Victoria, British Columbia, V8P 1A1, Canada\\
\scriptsize
$^{2}$Department of Physics, Kavli Institute for Astrophysics and Space Research, Massachusetts Institute of Technology, Cambridge, MA 02139, USA\\
\scriptsize
$^{3}$National Research Council of Canada, Herzberg Institute of Astrophysics, 5071 West Saanich Road, Victoria, British Columbia, V9E 2E7, Canada
}

\date{Accepted XXX. Received YYY; in original form ZZZ}

\pubyear{2016}

\begin{document}
\label{firstpage}
\pagerange{\pageref{firstpage}--\pageref{lastpage}}
\maketitle

\begin{abstract}
The interpretive power of the newest generation of large-volume hydrodynamical simulations of galaxy formation rests upon their ability to reproduce the observed properties of galaxies. In this second paper in a series, we employ bulge+disc decompositions of realistic dust-free galaxy images from the Illustris simulation in a consistent comparison with galaxies from the Sloan Digital Sky Survey (SDSS). Examining the size-luminosity relations of each sample, we find that galaxies in Illustris are roughly twice as large and $0.7$ magnitudes brighter on average than galaxies in the SDSS. The trend of increasing slope and decreasing normalization of size-luminosity as a function of bulge-fraction is qualitatively similar to observations. However, the size-luminosity relations of Illustris galaxies are quantitatively distinguished by higher normalizations and smaller slopes than for real galaxies. We show that this result is linked to a significant deficit of bulge-dominated galaxies in Illustris relative to the SDSS at stellar masses $\log\mathrm{M}_{\star}/\mathrm{M}_{\odot}\lesssim11$. We investigate this deficit by comparing bulge fraction estimates derived from photometry \emph{and} internal kinematics. We show that photometric bulge fractions are systematically lower than the kinematic fractions at low masses, but with increasingly good agreement as the stellar mass increases.
\end{abstract}
\begin{keywords}
galaxies: structure -- hydrodynamics -- surveys -- astrophysics
\end{keywords}



\section{Introduction}

The observed relationship between size and luminosity is a crucial benchmark within the framework of hierarchical assembly of galaxies \citep{1996ApJ...464L..63S,2003MNRAS.343..978S,2004ApJ...604..521T,2007ApJ...671..203C,2008MNRAS.388.1708G,2011ApJ...728...51B,2011MNRAS.410.1660D,2012MNRAS.421..621B,2014MNRAS.440L..51C,2015arXiv151005645F,2015MNRAS.450.1937C}. The morphologies that are determined by photometric analyses are governed by the growth and evolution of stellar populations and their distribution within galaxies. Reproducing the observed size-luminosity relation of galaxies within hydrodynamical simulations requires adequate numerical resolution and a broad physical model that includes key physical processes: stellar and gas kinematics; gas-cooling; star formation, feedback, and quenching; stellar population synthesis and evolution; black hole feedback, and the influence of galaxy interactions and merging on these processes \citep{2011ApJ...728...51B,2014MNRAS.440L..51C,2015arXiv151005645F,2015MNRAS.450.1937C}. The similarities and differences between the size-luminosity relations of the simulated and observed galaxies reflect the successes and trappings of the models employed by the simulations.

The sizes of galaxies in hydrodynamical simulations have only recently demonstrated consistency with observations. In particular, the formation of realistic disc galaxies in earlier generations of simulations was recognized as a significant challenge within a framework of hierarchical assembly (e.g., \citealt{1995MNRAS.275...56N,1997ApJ...478...13N}). Simulated discs were too centrally concentrated, too small, and rotated too quickly at fixed luminosity. The resulting Tully-Fisher relations and disc angular momenta in early disc-formation experiments yielded stark contrasts with observations (e.g., \citealt{1999ApJ...513..555S,2000ApJ...538..477N,2001ApJ...554..114E,2003ApJ...591..499A,2004ApJ...607..688G,piontek2009angular}; review by \citealt{2010ASPC..432...17B}; also see comparison of various hydrodynamical codes by \citealt{2012MNRAS.423.1726S}). The inclusion of energetic feedback has been shown to mitigate the differences between simulated and observed discs by preventing overcooling of gas, runaway star formation at early times, and angular momentum deficiency in simulated disc galaxies (e.g., \citealt{2010MNRAS.409.1541S,2011MNRAS.415.1051B,2012MNRAS.419..771B}). High-resolution hydrodynamical simulations that include efficient feedback have yielded more reasonable disc sizes in small galaxy samples and for targeted mass ranges \citep{2004ApJ...607..688G,2005MNRAS.363.1299O,2010MNRAS.409.1541S,2012MNRAS.424.1275B,2012MNRAS.420.2245M,2013ApJ...766...56M,2013MNRAS.434.3142A,2014MNRAS.445..581H,2014MNRAS.437.1750M}. However, reproducing the size-mass and size-luminosity relations for galaxy \emph{populations} remains challenging in cosmological simulations. The size-mass and size-luminosity relations of galaxies depend sensitively on stellar mass and luminosity functions, the $\mathrm{M}_{\star}-\mathrm{M}_{\mathrm{halo}}$ relations, and feedback models -- which must all be accurate to reproduce observed galaxy sizes \citep{2012MNRAS.423.1726S,2015MNRAS.450.1937C}.

Galaxy sizes in statistically meaningful samples from hydrodynamical simulations and their dependencies on sub-grid models for star-formation and energetic feedback have been studied in several recent works (e.g., OWLS $z=2$: \citealt{2010MNRAS.409.1541S}; OWLS $z=0$: \citealt{2012MNRAS.427..379M}; GIMIC $z=0$: \citealt{2009MNRAS.399.1773C}). The sub-grid feedback parameters generally have large uncertainties and are often calibrated to reproduce global scaling relations for galaxy populations at specific epochs (e.g., see \citealt{2014MNRAS.445..175G}). In the Illustris simulation, the parameters for the efficiency of energetic feedback are calibrated to roughly reproduce the history of cosmic star-formation rate density and the $z=0$ stellar mass function \citep{2013MNRAS.436.3031V}. However, the \emph{evolution} of these relations are predictions of the simulation. \cite{2015MNRAS.454.1886S} performed an image-based comparison using non-parametric morphologies derived from mock Sloan Digital Sky Survey (SDSS) observations of galaxies from the Illustris simulation \citep[T15]{2015MNRAS.447.2753T} to show that galaxies in Illustris were roughly twice the size of observed galaxies for the same stellar masses at $z=0$. In the \textsc{EAGLE} simulation \citep{2015MNRAS.446..521S}, the feedback efficiency parameters were calibrated to reproduce the galactic stellar mass function and the sizes of discs at $z=0$ and the observed relation between stellar mass and black-hole mass \citep{2015MNRAS.450.1937C}. \textsc{EAGLE} has been shown to successfully reproduce the \emph{evolution} of passive \emph{and} star-forming galaxy sizes out to $z=2$ using inferred scalings between \emph{physical} and photometric properties \citep{2015arXiv151005645F}. Comparison of the predictions from large-volume and high-fidelity hydrodynamical simulations such as Illustris and \textsc{EAGLE} to observations enables improved constraints on the sub-grid physics that govern galaxy sizes. However, it is important that such comparisons be made in a fair way by deriving the properties of galaxies consistently in observations and simulations. Realistic mock observations of simulated galaxies make it possible to perform a direct, image-based comparison between models and real data.

Creating mock observations of simulated galaxies is the most direct way to consistently derive galaxy properties for comparisons with observations -- as the same analysis tools can be used to derive the photometric and structural properties of each. Mock observations of galaxies from hydrodynamical simulations have been successful in reproducing observed trends for targeted morphologies -- but have been limited to small samples of galaxies. \cite{2011ApJ...728...51B} (see also \citealt{2011MNRAS.410.1391A}) used high-resolution zoom-in hydrodynamical cosmological simulations and dust-inclusive radiative transfer to produce mock observations of a sample of eight disc galaxies. Bulge+disc decompositions of the surface-brightness profiles were performed on $B-$band images to estimate the disc scale-length, $r_d$ (or often, $h$, in the literature) and magnitudes of the bulge and disc components. The size-luminosity relation of the discs agreed well with observed discs at redshifts $z=0$ and $z=1$ \citep{2000AJ....119.2757V,2008MNRAS.388.1708G,2003ApJ...582..689M,2008ApJ...680...70M,2011ApJ...741..115M}. In particular, the $z=0$ discs were consistent with the size-luminosity relations of the observational samples within as large of dynamic range in magnitude as for the observations.

The size-luminosity relation for the bulges within discs has also been compared with observational constraints using mock photometry. \cite{2014MNRAS.440L..51C} identified two galaxies with significant bulge components within the high-resolution disc galaxies simulated by \cite{2011ApJ...728...51B}. The properties of the bulge components were derived from $H$-band photometry using bulge+disc decompositions -- consistently with photometric decompositions of observed bulges in late-type galaxies and in ellipticals \citep{2010ApJ...716..942F,2013ApJ...764..174F}. The properties of the bulges identified in the simulations were in broad agreement with the size-luminosity relation derived from observations but within a more narrow magnitude range than previously shown for the discs and a significantly more limited sample size.

The principal limitations of previous studies aimed at comparing simulated and observed structural relations include: (1) small sample sizes; (2) inconsistent derivations of simulated and observed galaxy properties; (3) incomplete observational realism that biases the distributions of derived properties of simulated galaxies in comparisons with observations. Each of these limitations can be addressed using realistic mock-observations of galaxies from large-volume cosmological hydrodynamical simulations. The current generation of large-volume hydrodynamical simulations contain sizeable populations of galaxies which can be used to compare the distributions of galaxies and their morphologies on the global size-luminosity relation. Mock observations of galaxies from these simulations (e.g., \citealt{2015MNRAS.447.2753T,2015MNRAS.452.2879T}) and observational realism (\citealt{2015MNRAS.454.1886S,Bottrell2016}) ensure that their derived properties are affected by the same observational biases as real galaxies. 

In \cite{Bottrell2016} we detailed the design and implementation of a new methodology for performing image-based comparisons between galaxies from cosmological simulations and observational galaxy redshift surveys. Our goal was to remove prior limitations to consistent morphological comparisons between theory and observations. In addition to using mock images, we also applied observational realism to these mock images to ensure the \emph{same} biases were present in the mock and real data. In the first implementation of the methodology, we presented catalogs of parametric bulge+disc decompositions for $\sim7000$ galaxies in the $z\sim0$ Illustris simulation snap shot with SDSS realism and a technical characterization of the effects of observational biases on some key parameters. In particular, the catalogs enable consistent comparisons with the existing bulge+disc decomposition catalogs of \citealt{2011ApJS..196...11S} \citepalias{2011ApJS..196...11S} for 1.12 million galaxies in the SDSS.

The distribution of a population of galaxies on the size-luminosity plane is governed by the distribution of stellar populations within the physical components of galaxies. The size-luminosity relation is therefore well suited to examine the successes and discrepancies in the structural morphologies of galaxies from the Illustris simulation using the methods and catalogs from our previous paper. In this second paper in a series, we employed our mock and real galaxy structural parameter catalogs to make the comparison between Illustris and the Sloan Digital Sky Survey (SDSS) as fair as currently possible. A review of our methods and description of our samples are presented in Section \ref{sec:methods}. In Section \ref{sec:slrs}, we compare size-luminosity luminosity relations of Illustris and SDSS. In Section \ref{sec:bulgedisc}, we examine the impact of morphological differences between the observed and simulated galaxy populations on the size-luminosity relations. In Section \ref{sec:deficit}, we investigate the connection between morphology and stellar mass in SDSS and Illustris and compare the photometric and kinematic bulge-to-total fractions of galaxies from the Illustris simulation. We discuss and summarize our results in Sections \ref{sec:discussion} and \ref{sec:summary}, respectively.

\section{Methods}\label{sec:methods}

\subsection{Illustris simulation}

A detailed description of the Illustris simulation can be found in \cite{2014MNRAS.444.1518V}, \cite{2014Natur.509..177V}, and \cite{2014MNRAS.445..175G}. In this section, we briefly summarize the Illustris simulation properties that are most relevant to the creation of the synthetic images and our comparison.

Illustris is a cosmological hydrodynamical simulation that is run in a large cubic periodic volume of side-length $L = 106.5$ Mpc. The simulation is run using the moving-mesh code \textsc{arepo} \citep{2010MNRAS.401..791S} and a broad physical model that includes a sub-resolution inter-stellar medium (ISM), star-formation, and associated feedback \citep{2003MNRAS.339..289S}, gas cooling \citep{1996ApJS..105...19K,2009MNRAS.393...99W}, stellar evolution and enrichment \citep{2009MNRAS.399..574W}, heating and ionization by a UV background \citep{2008ApJ...681..831F,2009ApJ...703.1416F,2009ApJ...694..842M}, black hole seeding, merging, and active galactic nucleus (AGN) feedback \citep{2005MNRAS.361..776S,2006MNRAS.366..397S,2007MNRAS.380..877S,2014MNRAS.442.1992H}. Details of the physical model employed in Illustris can be found in \cite{2013MNRAS.436.3031V} and \cite{2014MNRAS.438.1985T}. The volume contains $N_{\mathrm{DM}}=1820^3$ dark matter particles ($m_{\mathrm{DM}}=6.3\times10^6\mathrm{M}_{\odot}$) and $N_{\mathrm{baryon}}=1820^3$ gas resolution elements ($m_{\mathrm{baryon}}\approx1.3\times10^6$). Stellar particles ($M_{\star}\approx1.3\times10^6 \mathrm{M}_{\odot}$) are formed stochastically out of cool, dense gas resolution elements and inherit the metallicity of the local ISM gas. Stellar particles then gradually return mass to the ISM to account for mass-loss from aging stellar populations. The age, birth mass, and time-dependent current mass are tracked for each stellar particle within the simulation. The gravitational softening lengths of dark and baryonic particles are $\epsilon_{\mathrm{DM}}=1420$ pc and $\epsilon_{\mathrm{baryon}}=710$ pc, respectively. The smallest gas resolution elements at $z=0$ have a typical extent (fiducial radius) $r_{\mathrm{cell}}^{\mathrm{min}}=48$ pc. Haloes are defined in the Illustris simulation using a Friends-of-Friends (FoF) algorithm (e.g., \citealt{1985ApJ...292..371D}) with a linking length of 0.2 times the mean particle separation to identify bound haloes. Individual galaxies are defined with the \textsc{subfind} halo-finder \citep{2001ApJ...549..681S}.

The initial conditions for the simulation assume a $\Lambda$CDM model consistent with WMAP-9 measurements \citep{2013ApJS..208...19H}: $\Omega_{\text{M}} = 0.2726$; $\Omega_{\Lambda} = 0.7274$; $\Omega_{\text{b}} = 0.0456$; $\sigma_8 = 0.809$; $n_s= 0.963$; and $H_0 = 100 h$ km s$^{-1}$Mpc$^{-1}$ where $h=0.704$. Free parameters within the Illustris model were calibrated in smaller simulation volumes to \emph{roughly} reproduce the observed galaxy stellar mass function at $z=0$ and star-formation rate density across cosmic time.

\cite{2014MNRAS.444.1518V}, \cite{2014Natur.509..177V}, and \cite{2014MNRAS.445..175G} examine the physical properties of galaxies from Illustris in a comparison with several key observations. The cosmic star-formation rate density and galaxy stellar mass functions agree reasonably well with observations, by construction. Still, Illustris produces slightly too many galaxies with masses $\log \mathrm{M}_{\star}/\mathrm{M}_{\odot}<10$ and $\log \mathrm{M}_{\star}/\mathrm{M}_{\odot}>11.5$ relative to observations \citep{2013MNRAS.436..697B,2013ApJ...767...50M} (though it must be noted that these observations have significant measurement uncertainties at the high-mass end). The cosmic star-formation rate density in Illustris, while accurately reproducing the observed trend between $z\sim1-10$ \citep{2013ApJ...770...57B,2013ApJ...763L...7E,2013ApJ...773...75O}, is also slightly too large at $z=0$ -- corresponding to larger fractions of star-forming/blue galaxies for stellar masses $\log \mathrm{M}_{\star}/\mathrm{M}_{\odot}<10.5$ \citep{2016MNRAS.462.2559B}. Nonetheless, the global passive/red and star-forming/blue fractions for galaxies with $\log \mathrm{M}_{\star}/\mathrm{M}_{\odot}>9$ at $z=0$ agrees reasonably well with observations (though colours become less accurate within specific stellar mass domains). 

The Illustris $r$-band galaxy luminosity function at $z=0$ also reasonably agrees with local observations from the SDSS for $M_r\sim-15$ to $-24$ \citep{2013MNRAS.436..697B} as shown by \citet{2014MNRAS.444.1518V}. Visualization of galaxies with over $10^5$ stellar particles, $\log \mathrm{M}_{\star}/\mathrm{M}_{\odot}\gtrsim11$, demonstrates that Illustris produces diverse morphological structures including populations of star-forming blue discs and passive red bulge-dominated galaxies. Kinematic bulge-to-total stellar mass fractions of these well-resolved galaxies were used to demonstrate that Illustris accurately describes the transition from late- to early-types as a function of total stellar mass -- finding reasonable agreement with observational \emph{photometric} morphological classifications presented in \cite{2006MNRAS.373.1389C}. \cite{2014MNRAS.444.1518V} cautions that the morphological comparison should not be over-interpreted because the morphologies from Illustris were classified physically, where the observed morphologies were classified visually. On the other hand, the methodology used in this paper is particularly well-suited to perform a consistent and detailed comparison of galaxy morphologies using the same classification methods. 

Detailed comparisons between Illustris and the observed galaxy scaling relations can be found in \citet{2014MNRAS.444.1518V} and \citet{2014MNRAS.445..175G}, which specifically examine the galaxy luminosity functions, stellar mass functions, star formation main sequence, Tully-Fisher relations, stellar-age stellar-mass relations, among others. Those papers show that Illustris broadly reproduces the redshift $z=0$ galaxy luminosity function, the evolving galaxy stellar mass function, and Tully-Fisher relations better than previous simulations that included less developed feedback models \citep[e.g.,][]{1994MNRAS.267..401N,2009MNRAS.399.1773C,2012MNRAS.425.3024V, 2014MNRAS.438.1985T}. However, there are some areas (e.g., the mass-metallicity relation, size-mass relation, or stellar-age stellar-mass relation) where significant tension remains between simulated and observed results. In this paper, we consider a more stringent test for comparing the Illustris simulation results against observations by applying even-handed analysis to synthetic Illustris observations, and real SDSS images. Our approach is aimed at identifying specific conflicts with the models and observations that can be used to refine future generations of galaxy formation models.

\subsection{Stellar mocks}\label{sec:stellar_mocks}
We employ mock observations of galaxies from the redshift $z=0$ snapshot of the Illustris simulation taken from the synthetic image catalog of \citealt{2015MNRAS.447.2753T} \citepalias{2015MNRAS.447.2753T}. Each synthetic image of a galaxy is centred on the \emph{galaxy's} gravitational potential minimum with field-of-view dimensions equal to 10 times the stellar half-mass radii $rhm_{\star}$ of the galaxy (i.e., using particles/cells defined by \textsc{subfind}). Stellar particles within the \emph{full FoF group} are each assigned a spectral energy distribution (SED) based on their mass, age, and metallicity values using the \textsc{starburst99} (SB99) single-age stellar population SED templates \citep{1999ApJS..123....3L,2005ApJ...621..695V,2010ApJS..189..309L}. The SEDs assume a Chabrier Initial Mass Function \citep[IMF]{2003PASP..115..763C}, as does the Illustris simulation itself. The images are produced using the \textsc{sunrise} radiative transfer code \citep{2006MNRAS.372....2J,2010MNRAS.403...17J} with four viewing angles for each galaxy. The viewing angles are oriented along the arms of a tetrahedron defined in the coordinates of the simulation volume (e.g., CAMERA 0 parallel to the positive z-axis of the simulation volume). Each pinhole camera is placed 50 Mpc away from the centre of the tetrahedron -- which is positioned at the gravitational potential minimum of the galaxy. The projection of the stellar light from a galaxy is, therefore, effectively randomly oriented with respect to the galaxy's rotation axis. The fiducial camera resolution is $256\times256$ pixels. $10^8$ photon packets are used in the Monte Carlo photon propagation scheme. The resulting mock position-wavelength data cube may then be convolved with an arbitrary transmission function and have its pixel resolution degraded to match the desired instrument. \citetalias{2015MNRAS.447.2753T} confirmed that the number of photons used in the propagation is sufficient such that the resulting synthetic images are well converged. Still, \citetalias{2015MNRAS.447.2753T} warn that caution should be exercised when examining the detailed structure of low-surface brightness features due to the residual Monte Carlo noise that can manifest as fluctuations in pixel-to-pixel intensity.

The synthetic images are created without the dust absorption/emission functionalities of the \textsc{sunrise} code. A truly comprehensive procedure for creating realistic images of galaxies from a cosmological simulation that can be compared with observations must include an accurate treatment of dust. In Section 2.2.3 of \cite{Bottrell2016}, we summarize the challenges (detailed in \citetalias{2015MNRAS.447.2753T}) of generating a proper treatment of dust for synthetic images of galaxies from simulations that do not resolve the complex structure of the interstellar medium on spatial scales required to properly model the dust distribution ($\ll1$ kpc). Indeed, the challenges extend beyond numerical convergence for dust-inclusive radiative transfer in \textsc{sunrise} that might be overcome at higher computational expense for a sub-sample of galaxies. We acknowledge the limitation that the lack of dust presents to our comparison with observations and reserve treatment of dust until such a time that the effects and uncertainties associated with (particular) dust models on the synthetic galaxy images are resolved. Nonetheless, our current comparisons will be valuable standards for future comparisons that employ comprehensive, dust-inclusive radiative transfer to create synthetic galaxy images. Indeed, the methodology presented in \citetalias{2015MNRAS.447.2753T} was designed to enable seamless integration of such dust models in the radiative transfer. Owing to the lack of dust in the synthetic images, we expect our optical luminosities and sizes to represent upper limits -- particularly for edge-on viewing angles of discs. 

The images used in this paper use SB99 SED templates without nebular emission line contributions or \textsc{H II} region modelling. \citetalias{2015MNRAS.447.2753T} examined a model that accounts for the impact of nebular emission and dust obscuration from unresolved birth clouds on the emergent SEDs from young stellar particles. Nebular emission from young stars can contribute substantially to the flux in certain broad-band filters. \citetalias{2015MNRAS.447.2753T} accounted for birth cloud emission/obscuration by replacing the SB99 emission of young stellar particles ($t_{\mathrm{age}}<10^7$ yr) with \textsc{mappings-III} model emission assuming partially obscured young stellar spectra \cite{2008ApJS..176..438G} with added contributions from \textsc{H II} regions \citep{2005ApJ...619..755D,2006ApJS..167..177D,2006ApJ...647..244D}. \citetalias{2015MNRAS.447.2753T} found that the spatial distribution of light in the resulting synthetic images were very similar to those that did not employ the nebular emission model. Therefore, \citetalias{2015MNRAS.447.2753T} opted not to include modelling of nebular emission in their fiducial (public) synthetic images in order to minimize post-processing uncertainties (as for the dust) while still creating sufficiently realistic images that comparisons can be drawn against observations. 

A stellar light distribution (SLD) scheme is required to map discretized stellar particles to continuous light distributions. We employ the fiducial adaptive $16^{\mathrm{th}}$ nearest-neighbour SLD scheme from the \citetalias{2015MNRAS.447.2753T} public release. In \cite{Bottrell2016}, we showed that estimates of size and luminosity for most galaxies are largely invariant to the choice of SLD scheme. However, systematics from internal segmentation are appreciable for galaxies with stellar half-mass radii $rhm_{\star} > 8$ kpc and total stellar masses $\log\mathrm{M}_{\star}/\mathrm{M}_{\odot}<11$ and are not alleviated by any choice of SLD scheme that were examined in \cite{Bottrell2016}. Additionally, large constant smoothing radii ($\sim$1 kpc) tend to produce galaxies with systematically smaller $(B/T)$ than in the fiducial scheme. Ultimately, we followed the philosophy of \citetalias{2015MNRAS.447.2753T} that no SLD scheme is more physically motivated than another, and the choice to use the fiducial scheme is motivated largely by its simplicity. 

The synthetic image catalog includes 6891 galaxies with stellar masses $\log \mathrm{M}_{\star}/\mathrm{M}_{\odot}>10$ corresponding to a $N_{\star}\gtrsim10^4$ stellar particle number cut. All synthetic images are artificially redshifted to $z=0.05$ and are convolved with SDSS $g$ and $r$ filters. The raw synthetic images include no observational realism or noise apart from some residual Monte Carlo noise that may manifest in pixel-to-pixel intensity fluctuations for low surface brightness features.

\subsection{Observational realism}\label{sec:realism}

To enable consistent comparisons with observations, galaxies from the simulation must be mock observed with the same realism that affect observations of real galaxies. Building on previous work (e.g., \citealt{2015MNRAS.454.1886S}), in \cite{Bottrell2016} we designed an extensive methodology for adding observational biases to the synthetic images. Our method is designed to achieve the same statistics for sky brightness, resolution, and crowding as galaxies in observational catalogs by assigning insertions into real image fields probabilistically based on the observed positional distribution of galaxies. Specifically, the synthetic images fluxes are convolved with the reconstructed point-spread function (PSF), have Poisson noise added, and are inserted into SDSS $g$ and $r$ band corrected images following the projected locations of galaxies from the bulge+disc decomposition catalog of \citetalias{2011ApJS..196...11S}. The realism procedure ensures that the biases on the decomposition model parameters from resolution, signal-to-noise, and crowding are statistically consistent for simulated and real galaxies.

The biases associated with the added realism on structural measurements are characterized in \cite{Bottrell2016} and summarized here. The dominant contribution to error in the measured parameters is internal segmentation in galaxies with clumps of locally bright features in otherwise diffuse surface brightness profiles (roughly characterized by stellar half-mass radii $rhm_{\star} > 8$ kpc and total stellar masses $\log\mathrm{M}_{\star}/\mathrm{M}_{\odot}<11$). Internal segmentation in galaxies with locally bright features occurs because a deblending procedure is required to separate external sources from the galaxy photometry. Unrealistic stellar light distributions can lead to parts of a galaxy-of-interest being confused as an external source by the deblending. In extreme situations, photometric analysis may be reduced to a fraction of the original galaxy surface brightness distribution -- leading to large systematic and random errors in both magnitude and size for particular galaxies, as well as spurious measurements of $(B/T)$. Our analysis showed that \emph{some degree} of internal segmentation occurs in roughly $30\%$ of galaxies from Illustris. However, for galaxies not affected by internal segmentation, we showed that magnitude and half-light radii were robust to observational biases. Random errors in $(B/T)$ were generally larger for galaxies in which the bulge and disc components are both appreciable, but is a trend that is qualitatively consistent with decompositions of analytic bulge+disc models in the SDSS (see \citealt{2014ApJS..210....3M}, Appendix B).

\subsection{Bulge+disc decompositions}\label{catalogs}
Bulge+disc decompositions were performed on the mock observations from Illustris with the surface-brightness decomposition software \textsc{gim2d} \citep{1998ASPC..145..108S,2002ApJS..142....1S,2011ApJS..196...11S}. As described in \cite{Bottrell2016}, we model every realization of a galaxy with a single-component pure \Sersic{} profile (free \Sersic{} index, $n_{pS}$) and a two-component bulge+disc decomposition model with fixed bulge \Sersic{} index, $n_b=4$, and exponential disc, $n_d=1$, profiles. We focus on the bulge+disc decomposition results in this paper. The following catalogs were defined by \cite{Bottrell2016} and are used again in this paper:

\begin{enumerate}
\item [\textbf{\texttt{DISTINCT} catalog}:] A single bulge+disc decomposition for all galaxies and each of four camera angles ($\sim 28,000$ decompositions). Each camera angle incarnation of a galaxy is inserted into the SDSS following Section \ref{sec:realism}. Decompositions from the \texttt{DISTINCT} catalog are employed in our comparisons between mock-observed and real galaxies.
\item [\textbf{\texttt{ASKA} catalog}:] Multiple decompositions of a representative Illustris galaxy (RIG) sample of 100 galaxies that uniformly sample the stellar half-mass radius and total stellar mass distribution of Illustris galaxies from \citetalias{2015MNRAS.447.2753T}. Selection of the RIG sample is described in \cite{Bottrell2016}. Each RIG is inserted into roughly 100 SDSS sky areas following \ref{sec:realism} and all four camera angle incarnations of a galaxy are fitted at each location (leading to $\sim40,000$ decompositions in the \texttt{ASKA} catalog). The \texttt{ASKA} catalog enables measurement of collective uncertainties from biases such as resolution, sky brightness, and crowding on median measurements from the distributions of structural parameters. 
\end{enumerate}

\subsection{Selection of an SDSS comparison sample}\label{sec:controls}

The catalog of 1.12 million quantitative morphologies of galaxies from the SDSS by \citetalias{2011ApJS..196...11S} represents a reservoir from which we can draw populations of galaxies for comparisons to the simulated galaxies in the \texttt{DISTINCT} catalog. We compare our results with the $n_b=4$, $n_d=1$ bulge+disc decomposition results from the \citetalias{2011ApJS..196...11S} catalogs. However, several important criteria must be met for the comparison to be fair. While the design of the \texttt{DISTINCT} catalog ensured that biases from crowding, resolution, and sky were consistent between mock and real galaxies, we recall that all of our galaxies are inserted into the SDSS at redshift $z=0.05$. One observational bias that we have therefore not explored is the robustness of our parameter estimates with the surface brightness degradation as a function of redshift. A criterion of the SDSS control sample that is consequently necessary is that the control galaxies are confined to some thin redshift range around $z=0.05$ so that any biases that arise from surface brightness improvements or degradation do not enter into the comparison. Such a criterion for the redshift of a galaxy further requires that the estimate of the redshift is accurate -- which requires the additional criterion that galaxies must have spectroscopically measured redshifts. We therefore impose the following criteria on the \citetalias{2011ApJS..196...11S} catalog:

\begin{enumerate}
\item[(1)] Galaxies are selected from the Spectroscopic Sample of the SDSS DR7 Legacy Survey ($\sim660,000$ galaxies)
\item[(2)] Galaxies are confined to the volume corresponding to the spectroscopic redshift range $0.04<z<0.06$ ($\sim68,000$ galaxies)
\end{enumerate}


Biases from volume incompleteness in the samples of simulated and real galaxies are removed by sampling the galaxies in the \texttt{DISTINCT} catalog to match the normalized stellar mass distribution of the SDSS over $0.04<z<0.06$ with a lower mass cutoff of $\log \mathrm{M}_{\star}/\mathrm{M}_{\odot}>10$. The latter criterion is imposed because it is the stellar mass lower limit of Illustris galaxies for which there are synthetic images \citep{2015MNRAS.447.2753T}. The SDSS stellar masses are derived from combined surface brightness profile model estimates and SED template fitting by \cite{2014ApJS..210....3M}. Galaxies are drawn with replacement from the 28,000 galaxies in the \texttt{DISTINCT} catalog using a Monte Carlo accept-reject scheme to match the stellar mass distribution of the SDSS sample. The stellar masses for Illustris galaxies are computed from the sum of stellar particle masses that belong to a galaxy as identified by \textsc{subfind}. The differing methodologies for computing the stellar masses may introduce biases in the stellar mass matching \citep{2013MNRAS.435...87M,2014A&A...571A..75M,2015MNRAS.446.1512H}. \citetalias{2015MNRAS.447.2753T} showed that photometric masses derived from the synthetic galaxy SEDs from Illustris were broadly similar to the \emph{idealized} \textsc{subfind} masses from the simulation (with some systematics identified therein). However, to obtain stellar mass estimates for simulated galaxies with synthetic photometry that have the same inherent biases and uncertainties as the observationally derived masses would require dust-inclusive radiative transfer -- which first requires an accurate model for dust. Therefore, we acknowledge that comparing the photometrically-derived observed galaxy masses with the \textsc{subfind} masses of simulated galaxies may bias components of our analysis that depend on stellar mass matching. The exact role of the mass matching biases may be characterized using future high-resolution simulations that are better equipped to generate dust-inclusive synthetic photometry. 

\section{Galaxy size-luminosity relations}\label{sec:slrs}

\begin{figure*}
	\includegraphics[width=0.49\linewidth]{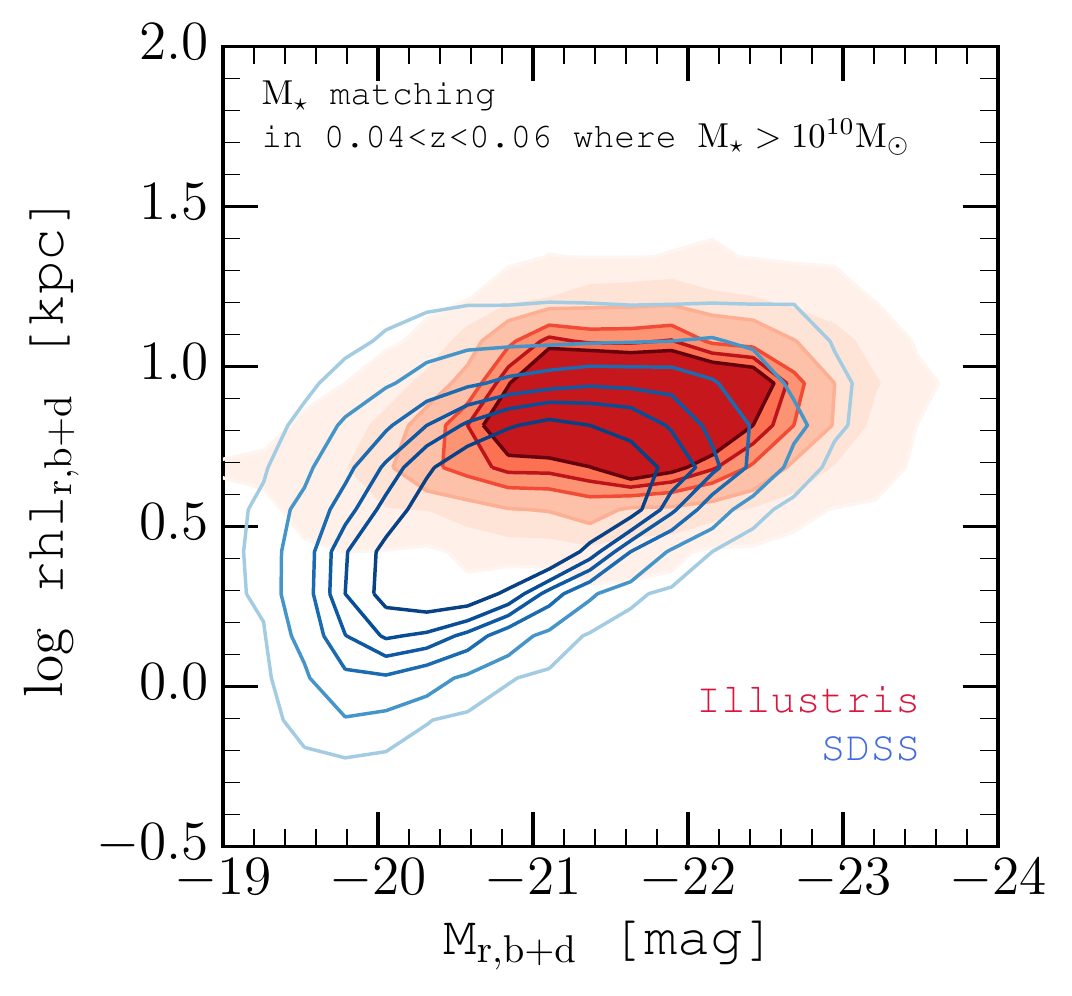}
	\includegraphics[width=0.49\linewidth]{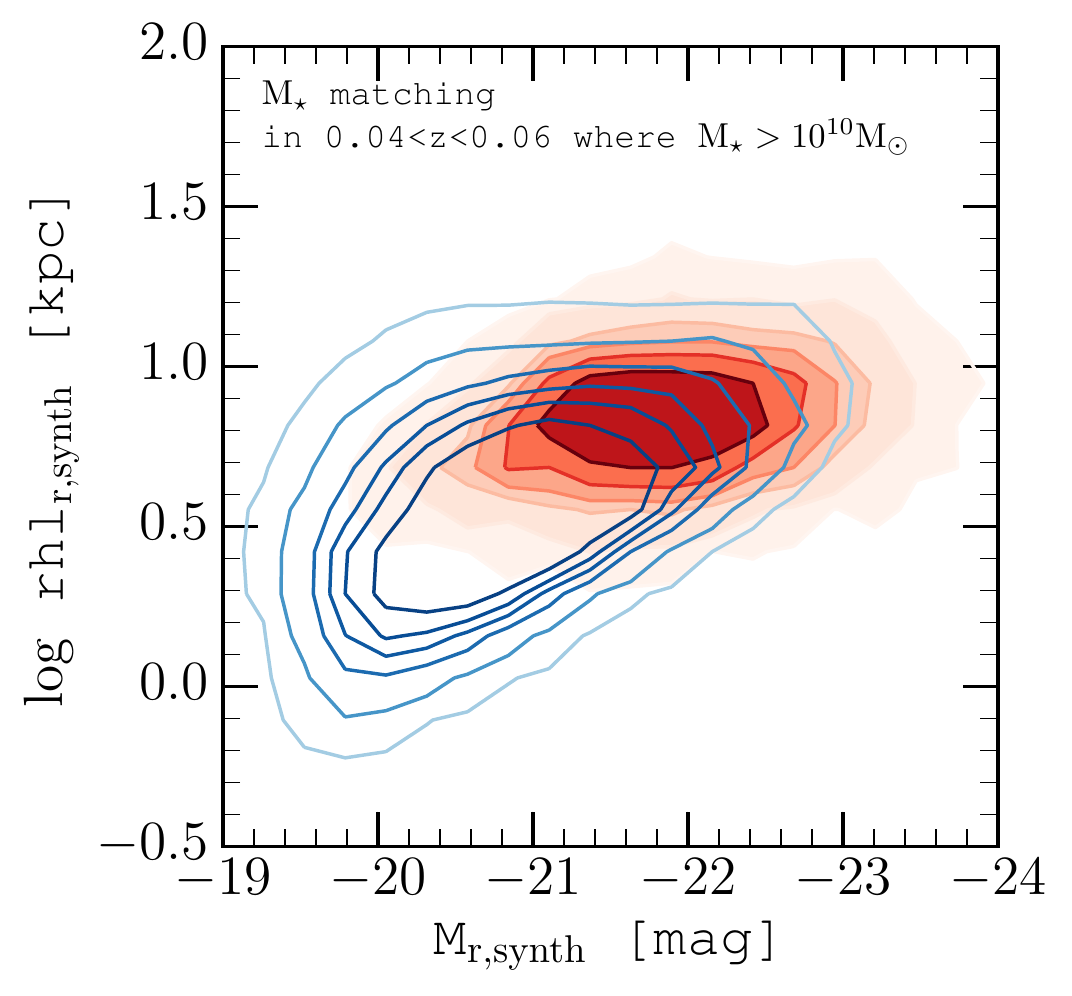}
    \caption[Size-luminosity relations of the SDSS and Illustris]{(\emph{Left}): The size-luminosity relations of the SDSS (blues) and Illustris (reds, filled) in samples matched by stellar mass. Illustris galaxy properties are taken from the \texttt{DISTINCT} catalog. Contour levels show the $50^{\mathrm{th}}$, $67^{\mathrm{th}}$, $75^{\mathrm{th}}$, $86^{\mathrm{th}}$, $95^{\mathrm{th}}$ and $99^{\mathrm{th}}$ percentiles for the size-luminosity distributions of Illustris and SDSS galaxies defined by $r$-band $b+d$ absolute magnitude and circular half-light radius in rest-frame physical coordinates. Illustris and SDSS each span roughly 4 magnitudes in luminosity, but galaxies in Illustris are intrinsically brighter and larger. The scaling between size and luminosity is also shallower for galaxies from Illustris -- leading to a stark contrast of size estimates at the low-luminosity end of the distributions. (\emph{Right}): Same as the left panel but using properties $M_{r,\mathrm{synth}}$ and $rhl_{r,\mathrm{synth}}$ computed directly from the flux in the synthetic images. The distribution for Illustris galaxies is shifted by roughly $0.2$ magnitudes brighter than in Figure \ref{fig:sizemag} due to the removal of bias from internal segmentation. Stellar masses for the SDSS galaxies are taken from the catalog of \citet{2014ApJS..210....3M}. $10^5$ Illustris galaxies are sampled with replacement to match the stellar mass distribution of the 34,700 galaxies in SDSS with $\log \mathrm{M}_{\star}/\mathrm{M}_{\odot}>10$ and are within $0.04<z<0.06$.}
\label{fig:sizemag}
\end{figure*}

The left panel of Figure \ref{fig:sizemag} shows the distributions of Illustris (red, filled contours) and SDSS (blue contours) in the plane of $r-$band half-light radius (as measured through circular aperture curve of growth photometry) and absolute $r-$band magnitude from the bulge+disc decompositions. The luminosities in each sample span roughly 4 magnitudes -- except for a low-luminosity tail in Illustris at the $99^{\mathrm{th}}$ percentile. However, the Illustris luminosities are brighter by roughly 0.7 magnitudes (factor of 2) on average. The left panel of Figure \ref{fig:sizemag} also demonstrates a discrepancy in sizes between the distributions of Illustris and SDSS galaxies. Galaxies with high luminosities, $M_r\lesssim-21.5$, are systematically larger in Illustris than galaxies observed in the real universe by roughly +0.4 dex (or a factor of 2 larger, consistent with \citealt{2015MNRAS.454.1886S}). There is also a discrepancy in the correlation between size and luminosity for Illustris galaxies with respect to the SDSS for the same stellar masses. The slope of the global size-luminosity relation for galaxies in Illustris is significantly shallower than for galaxies in the SDSS -- implying a weaker relationship between galaxy size and stellar mass in Illustris. 

The large offset in magnitude between Illustris and the SDSS in Figure \ref{fig:sizemag} occurs despite the known systematics from internal segmentation in an appreciable fraction of galaxies in the \texttt{DISTINCT} catalog \citep{Bottrell2016}. Roughly $30\%$ of galaxies in Illustris are affected by internal segmentation to some degree. The effect of internal segmentation on measured fluxes and sizes can be significant -- with reductions in total flux by \emph{up to} a factor of six \citep{Bottrell2016}. However, the offset in magnitude that is seen in Figure \ref{fig:sizemag} for size-luminosity distributions of Illustris relative to the SDSS galaxies is \emph{negative} -- opposite to the positive magnitude bias from internal segmentation. Indeed, the systematically larger magnitude estimates from internally segmented galaxies seem only to broaden the high-magnitude tail of the 99\% contour for Illustris galaxies. 


The right panel of Figure \ref{fig:sizemag} shows that replacing the decomposition results with galaxy properties derived directly from the synthetic images, $M_{\mathrm{r,synth}}$ and $rhl_{\mathrm{r,synth}}$, affects only the sizes and fluxes at low-luminosities, $M_{r,\mathrm{b+d}}\gtrsim{-20.5}$, and removes the low-luminosity outliers from Illustris. Ultimately, replacing the decomposition results with the synthetic image properties in the left panel of Figure \ref{fig:sizemag} generates an Illustris size-luminosity relation that is shifted by an additional $0.2$ magnitudes brighter with respect to the SDSS and no particular improvement to agreement in slope with SDSS. The biases from internal segmentation are therefore insufficient to explain the discrepant offset in magnitude and difference in slope in the size-luminosity relations of Illustris and the SDSS. Although informative on the effect of internal segmentation, it should be noted that the comparison in the right panel of Figure \ref{fig:sizemag} is biased. The quantities for the SDSS galaxies are derived from the decomposition models, while for the Illustris galaxies they are derived directly from the synthetic images. The differences in the Illustris size-luminosity distributions in the left and right panels cannot strictly be interpreted as arising from internal segmentation alone. However, in \cite{Bottrell2016}, we characterized the biases on half-light radii and magnitudes from various sources -- showing that in the absence of internal segmentation, half-light radii and magnitudes that are computed from the models and from the synthetic images are broadly consistent. 

Our dust-free synthetic images do not permit a characterization of the effects of dust in the differences in the size-luminosity distributions. The systematics from dust on model parameters for the bulge and disc differ -- complicating speculative arguments on how exactly global galaxy properties should be affected. In general, however, the optical luminosities shown here for Illustris should represent upper limits to the luminosities of a dusty galaxy population -- which is consistent with the shift to brighter magnitudes in Illustris. In Section \ref{sec:dust}, we discuss the role of dust in the context of bulge+disc decompositions of dusty galaxies and in our dust-free analysis.

\section{Impact of bulge and disc morphologies}\label{sec:bulgedisc}

\subsection{Morphological dependence of the size-luminosity relation}

Observational studies by \cite{2007ApJ...671..203C} and \cite{2010ApJ...716..942F} confirm differences between the size-luminosity relations of visually classified late-type and early-type galaxies. Discs are generally larger than bulges and samples that contain galaxies with dominant disc components are offset on the size-luminosity plane from samples of galaxies containing dominant bulges at fixed luminosity. The disc relation is also shallower and has more scatter than the size-luminosity relations of bulge-dominated galaxies. 

\cite{2010ApJ...716..942F} also performed a comparison of the size-luminosity relations for observed classical and pseudo-bulges -- finding that the size-luminosity relation of classical bulges is the same as for ellipticals. Pseudo-bulges, which are sometimes classified by \Sersic{} index, $n\lesssim2$, have a steeper slope than classical bulges and ellipticals on the size-luminosity relation but significantly greater scatter \citep{2009MNRAS.393.1531G}. The discrepancy between pseudo-bulges and classical bulges is expected as pseudo-bulges are believed to have a different formation mechanism and to be structurally different from classical bulges (e.g., \citealt{2004ARA&A..42..603K}, and references therein). Classification of bulges into pseudo-bulges and classical bulges using \Sersic{} index is generally imperfect \citep{2013pss6.book...91G}, but can be considered as an approximation \citep{2008AJ....136..773F}. Many bulges with $n<2$ follow the tight size-luminosity relation for classical bulges and, conversely, some bulges that are offset from this relation have \Sersic{} indices that are consistent with the distribution of classical bulges \citep{2009MNRAS.393.1531G,2013pss6.book...91G}. 

Our analysis does not discern between classical and pseudo-bulges -- as our bulge+disc decompositions use a bulge component with fixed \Sersic{} index, $n=4$. However, our analysis of the simulated and observed galaxy populations is internally consistent. If pseudo-bulges in the simulations are equally represented and structurally similar to observed pseudo-bulges, then their effect on the distribution of structural parameters will be the same. Therefore, while the structural estimates for pseudo-bulge properties may be inaccurate in our bulge+disc decompositions, any discrepancies between the distributions of bulge properties between the simulations and observations will be sourced by true structural differences of these components.


The presence and growth of a stellar bulge component in galaxy morphologies is strongly linked to many key processes of galaxy formation theory. The photometric bulge-to-total fractions obtained in the structural decompositions of galaxies provide estimates of the relative contribution of the bulge to their structure. The importance of bulges in various scaling relations including size-luminosity, the bulge-to-total fractions are well-suited to identify the morphological differences between Illustris and the SDSS. In this section, the observed size-luminosity relations of late-type (disc-dominated) and early-type (spheroid- or bulge-dominated) are examined to provide context for a morphological comparison using the photometric bulge-to-total fraction and total stellar mass. 

\begin{figure*}
	\includegraphics[width=0.49\linewidth]{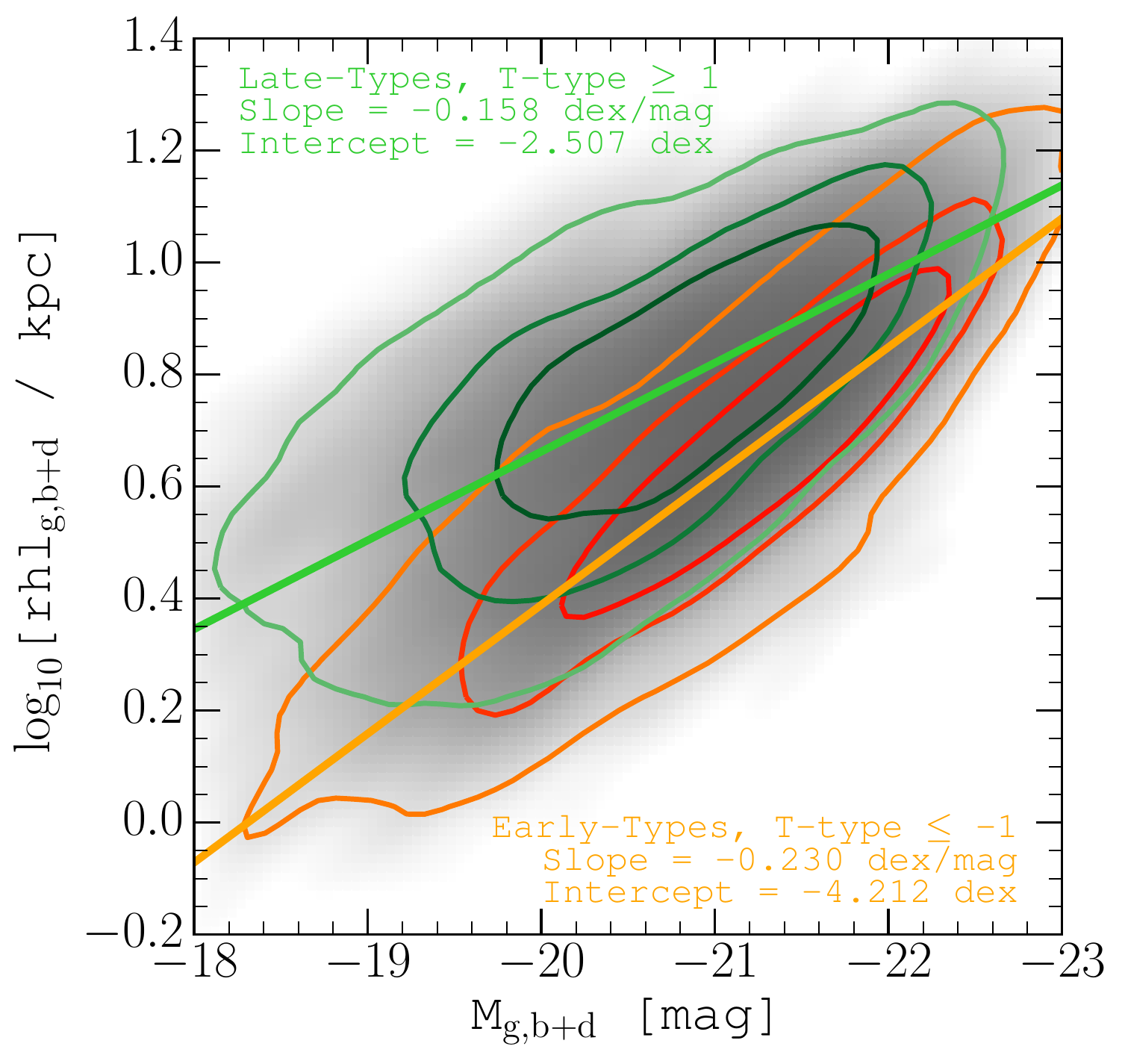}
	\includegraphics[width=0.49\linewidth]{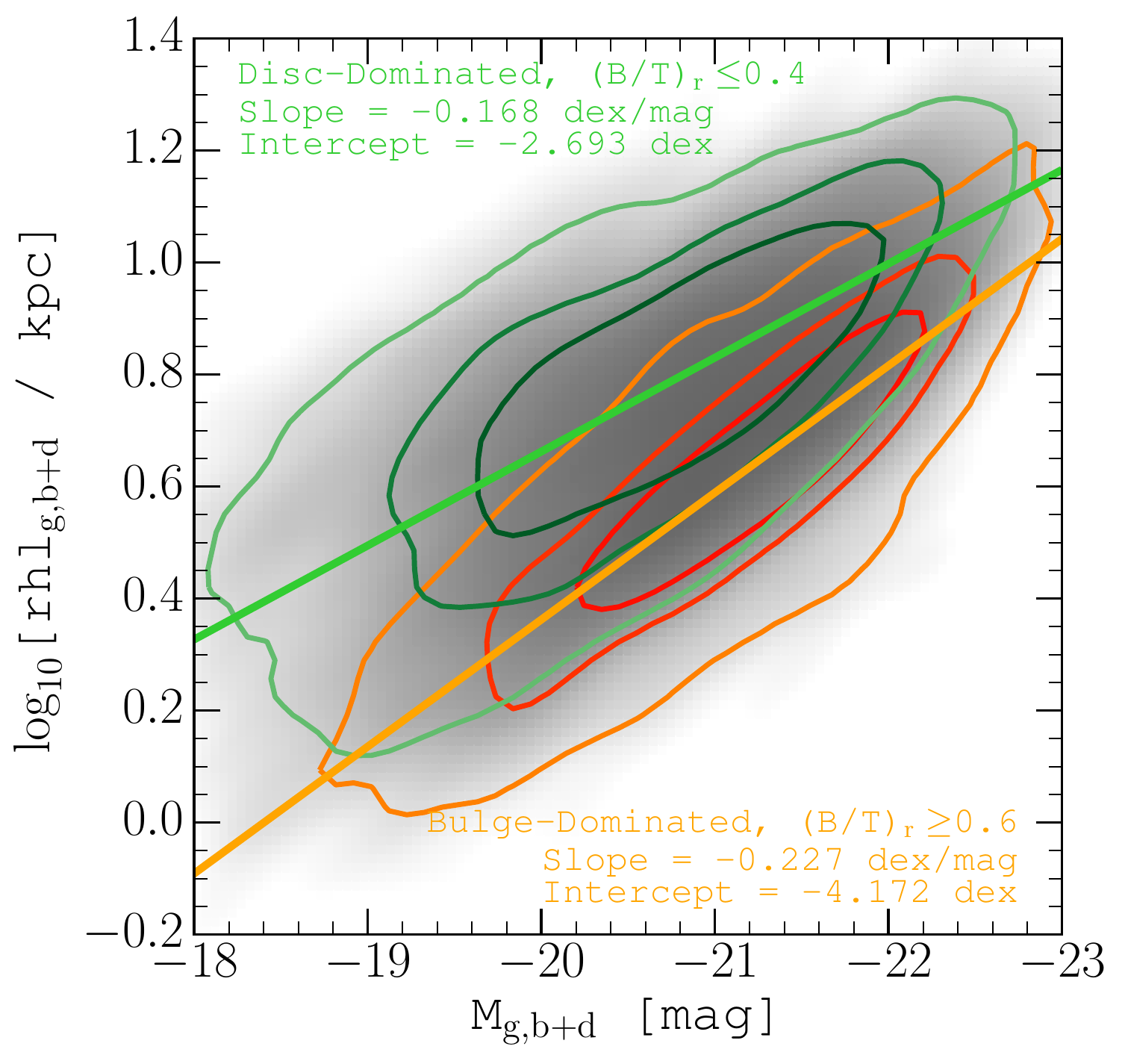}
    \caption[Size-luminosity relations of late- and early-type galaxies]{Size-luminosity relations of SDSS galaxies classified by visual and quantitative morphologies. Background shading in each panel shows the full distribution of 12,297 galaxies with $-18<M_g<-23$ taken from the visual classification sample of \cite{2010ApJS..186..427N} with magnitudes and half-light radii adopted from the bulge+disc decomposition catalog of \cite{2011ApJS..196...11S}. The left panel shows the size-luminosity relations for visually classified late- and early-type galaxies. Late-types (green) are selected with Hubble Type $T\geq1$ which roughly includes S(a,b,c,d) and Irregular galaxies. Early types (orange) are selected with $T\leq-1$ which roughly includes (c)E and S0 galaxies. The right panel shows the same sample of galaxies but split into disc-dominated $(B/T)\leq0.4$ and bulge-dominated $(B/T)\geq0.6$ systems using our quantitative morphologies. Contours show the $50^{\mathrm{th}}$, $75^{\mathrm{th}}$, and $95^{\mathrm{th}}$ percentiles in each classification group. The slope and scatter in the size-luminosity relations of late- and early-type galaxies are remarkably consistent between visual and quantitative splitting of the sample.}
    \label{fig:sizemag_bd}
\end{figure*}

\subsection{Bulge and disc fractions in Illustris and the SDSS}

The distinct size-luminosity relations of bulge and disk dominated galaxies is obvious in the SDSS sample, when populations are separated either by visual morphology, or quantitative bulge fractions. Figure \ref{fig:sizemag_bd} shows the size-luminosity relation of the visual classification sample of \cite{2010ApJS..186..427N} using the half-light radii $rhl_{\mathrm{g,B+D}}$ and absolute $g-$band magnitudes $M_{\mathrm{g,B+D}}$ from the bulge+disc decompositions of \cite{2011ApJS..196...11S}. The left panel of Figure \ref{fig:sizemag_bd} shows that roughly splitting the full sample into late- and early-types by Hubble $T$-type generates two distinct size-luminosity relations in slope and scatter. The result agrees well with other analyses of morphology dependence in the size-luminosity relations of galaxies (e.g., \citealt{2007ApJ...671..203C,2009ARA&A..47..159B}). The right panel of Figure \ref{fig:sizemag_bd} demonstrates that the morphological dependence of the galaxy size-luminosity relation is also seen when bulge-to-total fractions are used as a galaxy morphology indicator. The slope, scatter, and normalization of the size-luminosity relations of bulge- and disc-dominated systems separated by visual classification are almost exactly reproduced using bulge-to-total ratios from the quantitative bulge+disc decompositions. Therefore, $(B/T)$ estimates may enable sensitive investigation into the morphological differences between galaxies in the SDSS and Illustris that drive their size-luminosity relations. 

\begin{figure*}
	\includegraphics[width=1.01\linewidth]{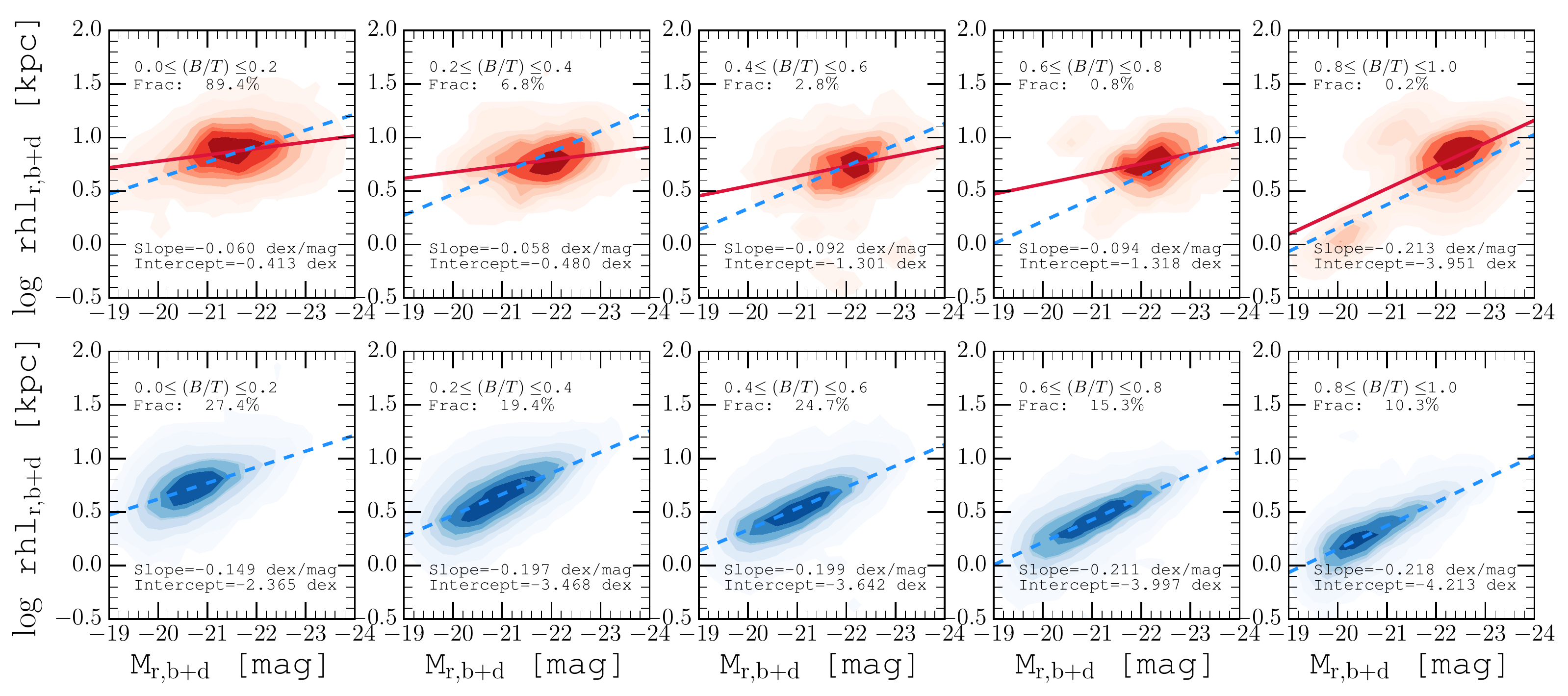}
    \caption[B/T morphology and the SLR]{Size-luminosity relations of Illustris (reds, upper panels) and the SDSS (blues, lower panels) for the mass matched galaxy samples from Figure \ref{fig:sizemag} with $(B/T)$ morphology classification (as before, $N_{\mathrm{Illustris}}=10^5$, $N_{\mathrm{SDSS}}=$34,700). Illustris galaxy properties are taken from the \texttt{DISTINCT} catalog. Contour levels show the $50^{\mathrm{th}}$, $67^{\mathrm{th}}$, $75^{\mathrm{th}}$, $86^{\mathrm{th}}$, $95^{\mathrm{th}}$ and $99^{\mathrm{th}}$ percentiles. In each row, panels are separated by $(B/T)$ classification -- indicated at the top left of each panel along with fraction of galaxies in each $(B/T)$ division compared to the whole. At the lower right of each panel, the slope and normalization of a linear fit to the distribution is also shown (black solid lines for Illustris, dashed for the SDSS). The best-fitting relation for SDSS in each $(B/T)$ interval is also plotted in the corresponding upper panel for visual impression.}
    \label{fig:sizemag_bt}
\end{figure*}

Figure \ref{fig:sizemag_bt} shows the size-luminosity relations of Illustris and the SDSS classified morphologically by $(B/T)_r$ using the same samples as for Figure \ref{fig:sizemag}. The Illustris size-luminosity relations demonstrate qualitatively similar changes with $(B/T)$ morphology as seen for the SDSS galaxies: the slope of the size-luminosity relation increases in higher $(B/T)$ classification groups and the normalization similarly decreases. However, though qualitatively similar, they are quantitatively distinct. Galaxies in Illustris have higher normalizations and shallower slopes across all $(B/T)$ classifications. Indeed, the fractions in each $(B/T)$ classification are also indicated -- forecasting a discrepancy in the morphological distributions of Illustris and the SDSS.

\begin{figure*}
	\center\includegraphics[width=\linewidth]{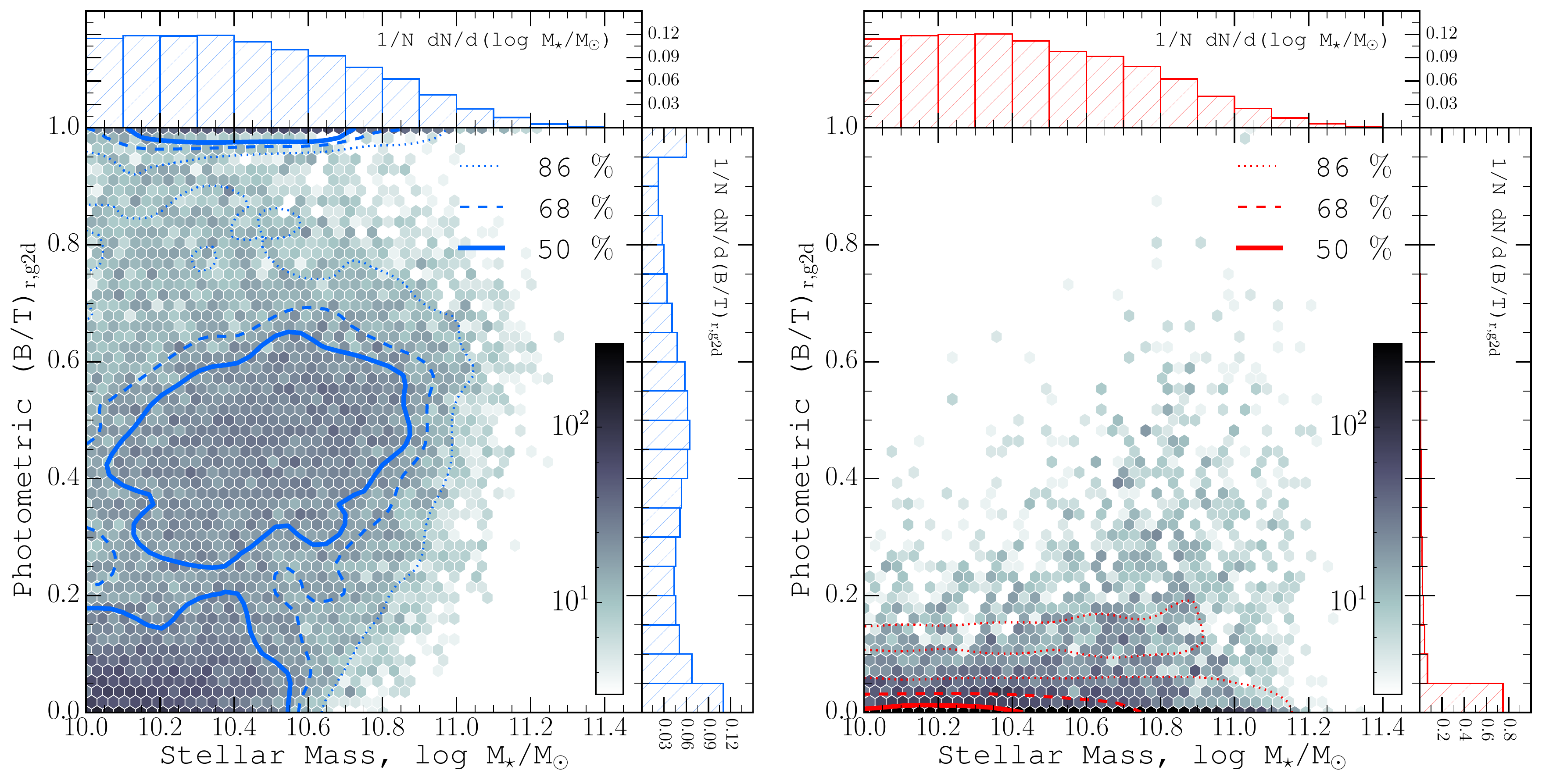}
    \caption[Bulge-to-total fractions in the SDSS and Illustris]{The distributions of galaxy bulge-to-total fractions in mass-matched galaxy samples of the SDSS (left panel, $N_{\mathrm{SDSS}}=$34,700) and Illustris (right panel, $N_{\mathrm{Illustris}}=10^5$) as a function of total stellar mass. The galaxies shown here are the same ones as in Figure \ref{fig:sizemag}. Illustris galaxy properties are taken from the \texttt{DISTINCT} catalog. Stellar masses for SDSS galaxies are taken from \cite{2014ApJS..210....3M} and are the same masses used in the matching of the samples. The upper marginals show the stellar mass distributions in each sample and demonstrate their consistency. The right marginals show the distributions of photometric bulge-to-total fractions, $(B/T)_r$. The central panels show the 2-dimensional distribution of total stellar mass and photometric $B/T$ on a logarithmic colourmap shown at along the lower right of the main panels. Coloured lines show the 50\% (solid), 68\% (dashed), and 86\% (dotted) contours of the distribution. The observations from the SDSS contains diverse morphologies classified by $B/T$ including galaxies dominated by both discs and bulges as well as a population of composite systems. Illustris galaxies are bereft of bulges in the corresponding mass-matched sample.}
    \label{fig:btr_mass}
\end{figure*}

Given that morphology is clearly critical in driving the normalization, slope, and scatter of the size luminosity relation, it is germane to compare the $(B/T)$ distributions of the SDSS and Illustris samples. Figure \ref{fig:btr_mass} shows the distribution of $r$-band photometric $(B/T)$ as a function of total stellar mass in the SDSS (left panel) and Illustris (right panel) taken from the \texttt{DISTINCT} catalog. The samples are the same as for Figure \ref{fig:sizemag} which were matched in stellar mass. Figure \ref{fig:btr_mass} shows that the SDSS has a diversity of morphological populations including bulge-dominated, disc-dominated, and a large number of composite galaxies within this mass distribution. The diversity is not shared by Illustris -- which is lacking in bulge dominated galaxies, particularly at low masses. Only for stellar masses $\log \mathrm{M}_{\star}/\mathrm{M}_{\odot}\gtrsim10.6$ do galaxies with significant bulge components become more common -- indicating a stronger correlation between bulge fraction and total stellar mass within Illustris than exists in the observations. The right marginal for the Illustris distribution shows that roughly 72\% of galaxies in Illustris have $(B/T)_r<0.05$, a fraction which rapidly declines for larger photometric bulge fractions (less than $1\%$ of galaxies in the Illustris sample have $(B/T)_r>0.6$).

The difference in morphological distributions between Illustris and SDSS shown in Figure \ref{fig:btr_mass} has an obvious implication for a comparison of their size luminosity relations. Illustris contains a much larger disc population than the SDSS for the sample matched by total stellar mass to the distribution of spectroscopic SDSS galaxies in $0.04<z<0.06$. Figure \ref{fig:sizemag_bd} showed that disc-dominated galaxies are elevated in the size-luminosity relation and have a shallower slope than early-types. The SDSS size-luminosity distribution in the comparison with Illustris in Figure \ref{fig:sizemag} is analogous to the background distributions for SDSS from Figure \ref{fig:sizemag_bd} -- showing the contributions of bulge- and disc-dominated galaxies to the relations. The galaxy size-luminosity relation for the SDSS is broadened vertically at low luminosities and has bent contours because it contains populations of discs, bulges, and composite systems. The bulges in SDSS weight the distribution to more compact half-light radii at low luminosities, as shown in Figure \ref{fig:sizemag_bd}. However, Illustris is deficient of bulges at low stellar masses (luminosities). Therefore, there is no downward weight from bulge-dominated galaxies at the low-luminosity end of Illustris to bring the slope and scatter of the galaxy size-luminosity relation into agreement with the SDSS. 

The impact of morphological differences between the SDSS and Illustris on the size-luminosity relation can be determined by matching samples in \emph{both} total stellar mass and bulge-to-total ratio. If the infrequency of bulge-dominated morphologies at low stellar masses in Illustris is responsible for the discrepancy in the size-luminosity relations of Illustris and the SDSS, then matching the SDSS morphologies (which are more diverse) \emph{and} stellar masses to the Illustris galaxies from Figure \ref{fig:sizemag} should bring the size-luminosity relations into better agreement.\footnote{Note that matching in the other direction does not work for SDSS galaxies in $0.04<z<0.06$. Illustris contains too few galaxies with high $(B/T)$ to be matched to the much larger population of bulges in the SDSS at these masses. In order to maintain the same stellar mass distributions as in our previous comparisons, we match galaxies one-to-one from the SDSS by stellar mass and $(B/T)$ to the sample of Illustris galaxies from Figures \ref{fig:sizemag} and \ref{fig:btr_mass} that are matched to the distribution of stellar masses in SDSS in $0.04<z<0.06$.} 

\begin{figure}
	\center\includegraphics[width=1.04\linewidth]{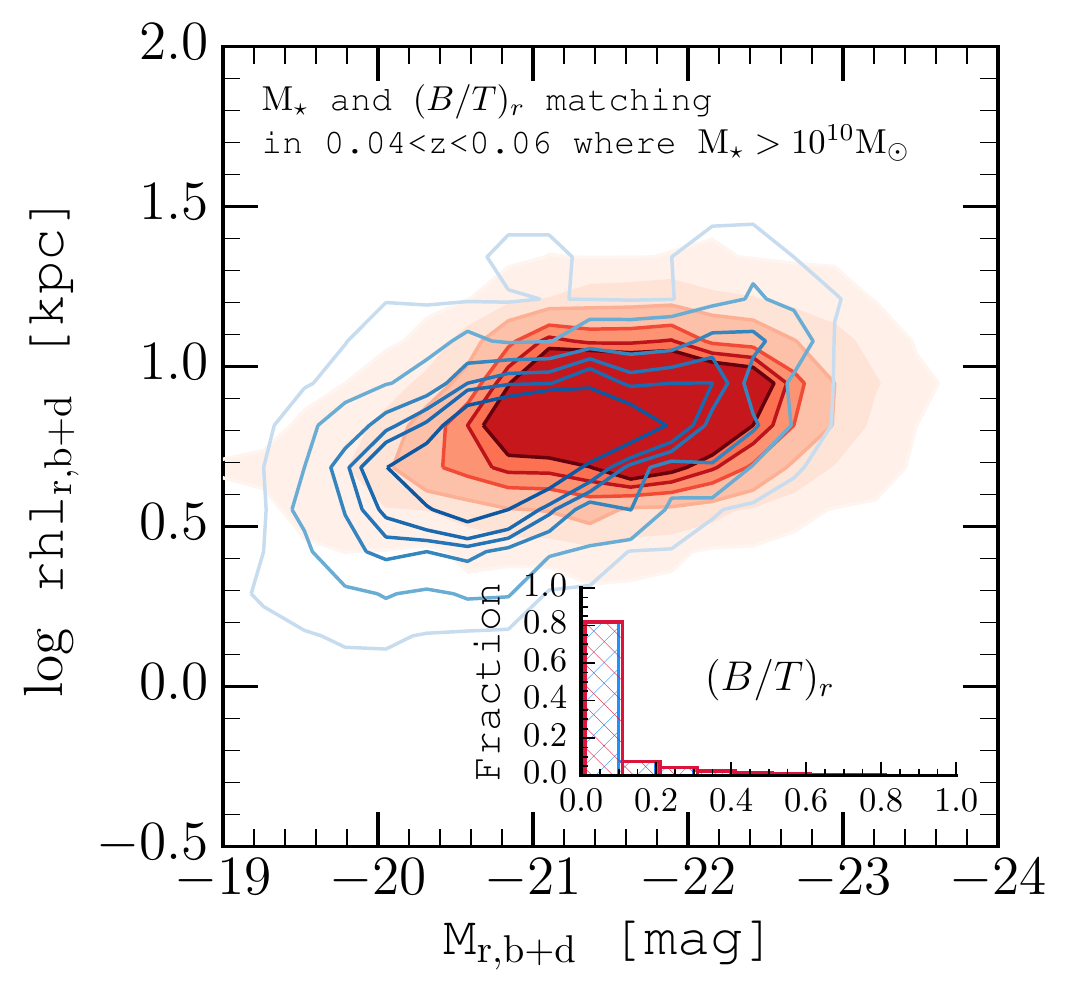}
    \caption[Size-luminosity relations matched by mass and morphology]{SDSS galaxies resampled to match the stellar mass \emph{and} $(B/T)$ morphology distributions of the Illustris galaxies from the left panel of Figure \ref{fig:sizemag} ($N_{\mathrm{Illustris}}=10^5$, $N_{\mathrm{SDSS}}=10^5$). Inset in the figure are histograms of $(B/T)_r$ for Illustris (red, hatched) and SDSS (blue, hatched opposite, offset for visual impression). Matching by morphologies significantly improves the agreement between the size-luminosity distributions. } 
    \label{fig:sizemag_btmatch}
\end{figure}

Figure \ref{fig:sizemag_btmatch} shows the size-luminosity relations for the SDSS and Illustris for samples now matched in both stellar mass and bulge-to-total fraction. The scaling and normalization between size and luminosity in Figure \ref{fig:sizemag_btmatch} for Illustris and SDSS galaxies is brought into greater agreement by the morphology matching. At low luminosities, the large discrepancy in galaxy sizes at fixed luminosity from Figure \ref{fig:sizemag} is largely removed. The improved agreement at low-luminosities is consistent with a deficit in bulges in Illustris relative to galaxies in the SDSS -- as seen in Figure \ref{fig:btr_mass}. Matching samples by morphology largely removes the bulge-dominated systems in the SDSS -- effectively leaving the size-luminosity relation of the discs. Still, the remaining disagreement indicates that, while the bulge deficit in Illustris may play a significant role in driving contrast with the SDSS size-luminosity relation, the morphology matching alone cannot provide a complete explanation for the differences in the size-luminosity relations of SDSS and Illustris. 

The remaining disagreement in the average sizes and luminosities of the mass-morphology matched samples could be due the lack of dust in the synthetic images. Dust corrections in synthetic images of galaxies have recently been considered for galaxies in the \textsc{EAGLE} simulation \citep{2015MNRAS.452.2879T}. Proper treatment and inclusion of dust effects might yield greater consistency in the observational realism of mock observations of galaxies.

\section{The deficit of bulges in low mass Illustris galaxies}\label{sec:deficit}

The results from the previous section demand further investigation into the lack of bulge-dominated galaxies in Illustris as determined by \textsc{gim2d} decompositions. The result contrasts with past problems in hydrodynamical simulations -- in which the \emph{physical parameters} of galaxies indicated that they were too bulge-dominated \citep{1994MNRAS.267..401N,1991ApJ...377..365K,2012MNRAS.423.1726S}. The question is whether (a) photometric bulges (photo-bulges), as identified by \textsc{gim2d}, systematically do not exist in Illustris; or (b) photo-bulges only do not strongly appear in the mass-matched sample that was used to examine the size-luminosity relation; or (c) true kinematic bulges are just not well extracted by \textsc{gim2d}. However, in the mass matched sample to $0.04<z<0.06$ of the SDSS, galaxies at the high mass end ($\log \mathrm{M}_{\star}/\mathrm{M}_{\odot}\gtrsim11$) of the $z=0$ stellar mass function of Illustris are not represented -- despite the larger cosmic volume in the observed sample.\footnote{There are a number of reasons for the larger number of high-mass galaxies in Illustris relative to the SDSS volume to which we performed the matching. One possible reason is that Illustris slightly over-predicts the redshift $z=0$ stellar mass function (SMF) at the high-mass end \citep{2014MNRAS.445..175G}. Further biases could arise in the analysis of galaxies at the centres of rich clusters -- which may provide discrepant stellar mass estimates with respect to the known masses from the simulations. A dedicated study of the systematics on photometric stellar mass estimates for all morphologies is required to fully understand the biases in the mass matching.} It is possible that the higher-mass galaxies have significant bulge components and are being missed in our analysis due to the standard of consistency we aim to achieve by matching in mass and redshift. Examination of the bulge fractions at higher masses in Illustris and comparisons with observations will yield insights on the discrepancy between the morphological dependence on stellar mass in Illustris and the SDSS.

In this section, the relationships between morphological $(B/T)$ fractions and stellar mass in the full populations of Illustris and SDSS are examined to provide insight on the deficit of bulges at low stellar mass in Illustris seen in the previous section. We argue possible scenarios that cause the discrepancies with observations. A comparison of the kinematically defined stellar bulge-to-total fraction in the simulation with the photometric fraction is also performed to examine the consistency between the information taken from the stellar orbits and stellar light.

\subsection{Morphological dependence on stellar mass}

\begin{figure*}
	\includegraphics[width=\linewidth]{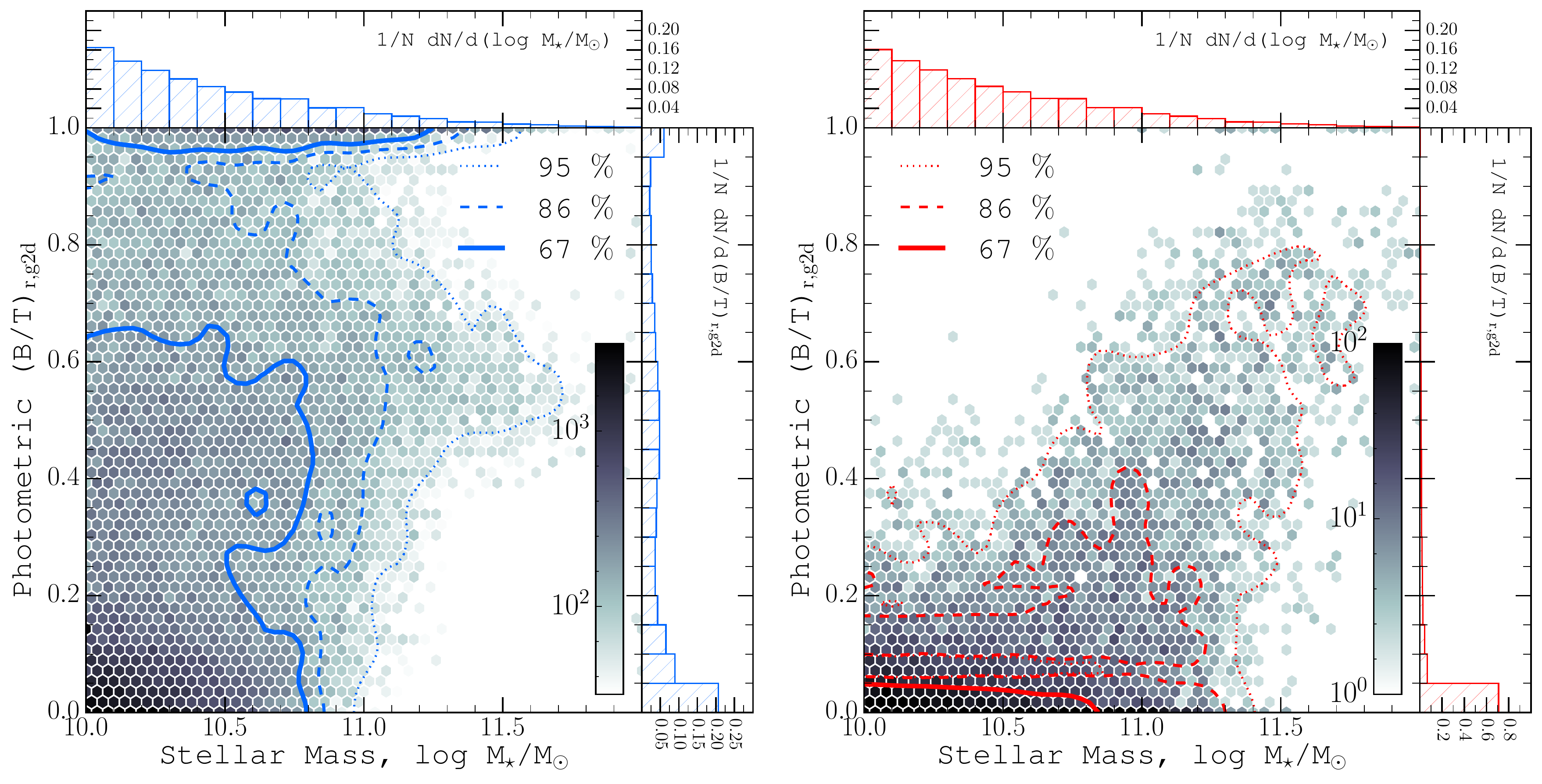}

    \caption[Mass dependence of $(B/T)$ in SDSS and Illustris]{SDSS (left panel) and Illustris (right panel) $r-$band bulge-to-total fraction distributions in samples matched to the stellar mass distribution of Illustris galaxies in the \texttt{DISTINCT} catalog. Each galaxy in the \texttt{DISTINCT} catalog is matched by stellar mass to 15 controls from the SDSS spectroscopic galaxy sample where $\log \mathrm{M}_{\star}/\mathrm{M}_{\odot}>10$ and $z<0.2$. Unlike Figure \ref{fig:btr_mass}, galaxies of all stellar masses from Illustris are represented in this comparison to show the stellar mass dependence of bulge-to-total fraction in Illustris ($N_{\mathrm{Illustris}}=$27,564, $N_{\mathrm{SDSS}}=$27,564 $\times$ 15).}
    \label{fig:sdss_illustris_bulges}
\end{figure*}

Galaxies in Illustris contain bulges -- albeit few at low stellar masses. Figure \ref{fig:sdss_illustris_bulges} shows the distribution of bulge-to-total fractions in the SDSS and Illustris with matching to the stellar mass distribution of Illustris galaxies from the \texttt{DISTINCT} catalog. We mitigate statistical biases by matching each galaxy from the \texttt{DISTINCT} catalog by stellar mass to the 15 nearest-mass neighbours in the SDSS $z<0.2$ control pool. We select from the $z<0.2$ SDSS volume to access galaxies that can be matched to the high-mass end of the Illustris stellar mass distribution -- for which there are too few in $0.04<z<0.06$. Taking galaxies from $z<0.2$ of SDSS will mean that spatial resolution biases are not controlled in this comparison -- though it is apparent that the SDSS distribution does not differ substantially from Figure \ref{fig:btr_mass}. The right panel of Figure \ref{fig:sdss_illustris_bulges} shows that there is a stronger relationship between $(B/T)$ and stellar mass in Illustris. At stellar masses $\log \mathrm{M}_{\star}/\mathrm{M}_{\odot} >11$ there appears to be a significantly more visible correlation between $(B/T)$ and stellar mass in Illustris than in SDSS -- where all disc, bulge, and composite morphologies are more evenly distributed as a function of stellar mass.

In Figure \ref{fig:sdss_illustris_bulges}, over 80\% of Illustris galaxies are completely disc-dominated at $\log \mathrm{M}_{\star}/\mathrm{M}_{\odot} < 10.5$. At slightly higher masses, $10.5<\log \mathrm{M}_{\star}/\mathrm{M}_{\odot}<11$, the SDSS have fewer disc-dominated systems and more composites, but Illustris still contains $\sim50\%$ discs with $(B/T)<0.05$ and few composites or bulge-dominated systems. The distributions become more similar in $11<\log \mathrm{M}_{\star}/\mathrm{M}_{\odot}<11.5$. Illustris contains an appreciable number galaxies with higher bulge fractions in $11<\log \mathrm{M}_{\star}/\mathrm{M}_{\odot}<11.5$ which is similar to the observations.  

Figure \ref{fig:sdss_illustris_bulges} demonstrates that although both simulated and real galaxies have a $(B/T)$-stellar mass dependence, there is a stronger correlation between photometric bulge-to-total ratio and total stellar mass in Illustris than in observed galaxies from the SDSS. Galaxies in the SDSS have diverse morphologies within each mass division -- whereas $(B/T)$ morphologies in galaxies from Illustris are strongly dependent on total stellar mass. Both populations share the trend that bulges become more frequent at higher stellar masses, but the dependence is stronger in Illustris. We now discuss possible biases that may explain these differences.



\subsection{Impact of stellar particle resolution/smoothing}

Accurate photometry for the inner region of the surface brightness distribution of a galaxy is essential for interpreting the bulge component \citep{2008MNRAS.388.1708G,2010MNRAS.403.2053G}. In \cite{Bottrell2016}, we showed that the choice of stellar light distribution (SLD) scheme did not bias the global properties of the galaxy such as total integrated magnitude and half-light radius, but could strongly bias the structural parameter $(B/T)$. We showed that broader smoothing kernels (such as the constant 1 kpc kernel relative to the $N=16$ nearest-neighbour kernel) artificially limit the spatial resolution in the inner regions of galaxies. Broad, constant smoothing kernels reduced concentration of flux from the central regions of the galaxy and systematically reduced estimates of $(B/T)$ systematically (to zero in many cases, even for galaxies with $(B/T)$ as large as 0.6 in the fiducial scheme). While it is possible that the fiducial SLD scheme may be biasing $(B/T)$ towards smaller bulge fractions in our decompositions, that leads to the notion that there is a ``correct'' SLD scheme for the particle mass resolution of Illustris (and the particle resolution of any hydrodynamical simulation). Some SLD schemes will give a better physical representation than others. But ultimately, the upper limit to the spatial resolution that is accessible through any SLD scheme is set by the stellar particle mass/spatial resolution in the simulation.

The choice of the $N_{16}$ nearest-neighbour smoothing as the fiducial model was motivated by simplicity \citep{2015MNRAS.447.2753T}. However, the comparisons of SLD schemes with narrower and broader smoothing kernels and their effects on the measured $(B/T)$ are qualitatively analogous to a comparisons using higher and lower stellar particle resolution, respectively. Higher particle mass resolutions (lower total stellar mass/particle) tend to reduce the typical spatial separation between stellar particles in galaxies produced in hydrodynamical simulations and effectively increase the spatial resolution (at least when using adaptive SLD schemes). Figure \ref{fig:sdss_illustris_bulges} showed that the majority of high mass systems ($\log \mathrm{M}_{\star}/\mathrm{M}_{\odot}\gtrsim11$) in Illustris (that contain larger numbers of particles) have bulges and that low mass galaxies largely do not. The spatial distribution of particles determines the surface brightness distribution of a simulated galaxy. Larger numbers of particles reduce the smoothing radii in our fiducial SLD scheme and generally improve the spatial resolution of a galaxy surface brightness distribution. Improvements to the spatial resolution in the bulge surface brightness distribution (in particular to the inner 100 pc that are essential for discerning its profile from a disc) may facilitate greater accuracy in modelling of the bulge component. If so, the strong mass dependence for the bulge-to-total fraction in Illustris seen in Figure \ref{fig:sdss_illustris_bulges} may arise from inadequate particle resolution for creating realistic photo-bulges in synthetic images of low mass galaxies with smaller numbers of stellar particles. 


One way to test the particle resolution dependence on bulge fractions directly is to perform hydrodynamical zoom-in simulations of lower mass systems in Illustris with the same numerical techniques and simulations models (e.g., for Illustris: \citealt{2016arXiv160408205S}). Comparison of the decomposition results from the high-resolution and low-resolution simulations would yield insight on the effects of particle resolution on $(B/T)$ estimates from mock observations. Alternatively, an investigation of the biases on structural morphology from the particle resolution and the simulation models that regulate the formation of structure may be performed by comparing our decomposition results with consistent decompositions of galaxies from other large hydrodynamical simulations such as \textsc{EAGLE} which has comparable mass resolution \citep{2015MNRAS.446..521S,2015MNRAS.450.1937C}. However, in such a comparison, the biases from differences in the simulation models on the morphological estimates would need to be carefully examined in order to assess whether particle resolution is the main culprit of the strong mass-dependence on $(B/T)$ estimates in the mock observations. Investigating simulations with different resolutions is beyond the scope of this paper.


\subsection{Comparison with kinematic B/T}\label{phot_vs_kin}


An interesting test that is feasible from our image-based decompositions of simulated galaxies is a comparison between the properties derived from kinematic and photometric information for the stellar particles. Comparisons of photometric and kinematic bulge fractions in galaxies have been performed previously in the literature without the large numbers or extensive realism considerations provided here. \cite{2010MNRAS.407L..41S} used bulge+disc decompositions of mock observations of Milky Way-mass galaxies from hydrodynamical simulations with similar mass resolution to Illustris ($\mathrm{M}_{\star}\sim10^6$) to show that $(B/T)$ is systematically lower from photometry relative to the kinematics. \cite{2016MNRAS.459..467O} reproduced this result in a sample of 18 cosmological zoom-in simulations of galaxies. Each galaxy from \cite{2016MNRAS.459..467O} had a photometric bulge-to-total ratio of $(B/T)\approx0$ but kinematic ratios $\sim0.5$. However, the zoom-in simulations from \cite{2016MNRAS.459..467O} used adaptive particle mass resolution to each halo -- making it difficult to ascertain the effects of particle mass resolution. Still, the implication of each study is that exponential structure of mock observed surface brightness profiles of simulated galaxies does not imply a cold rotationally-supported kinematic disc (i.e. low photometric $B/T$ does not necessarily indicate the lack of a kinematic bulge). These results are further complicated by \cite{2014MNRAS.440L..51C} who, while not specifically investigating the differences between photometry and the kinematics, demonstrated that realistic mock-observed photometric bulges can be produced in high-resolution simulations that match well with the photometric properties of real bulges. The implication from these studies is that while galaxies with realistic photo-bulges may be produced, bulge fractions inferred from photometry may not straightforwardly couple with kinematic bulge classifications.

In the simulations, the angular momentum for each particle about the principle rotational axis in a galaxy can be derived using the particle velocities and locations relative to the galactic potential. Stars that belong to the bulges of galaxies tend to have a gaussian distribution of velocities (see \citealt{2013ARA&A..51..511K} for a recent review) whereas stars in the disc have rotationally supported orbits and generally larger, coherent angular momenta. Stars (and particles representing stars) can be approximately associated to their stellar components using this angular momentum information to estimate the kinematic bulge-to-total or disc-to-total  fractions (e.g., \citealt{2009MNRAS.396..696S,2010MNRAS.408..812S}). One definition of the kinematic $(B/T)$ that is common in the literature (see \citealt{2003ApJ...591..499A,2015A&C....13...12N,2015ApJ...804L..40G}) is:
\begin{align}\label{eq:bt_kin}
(B/T)_{\mathrm{kin}} = 1/N_{\star,\mathrm{tot}} \quad 2 \times N_{\star}(J_z / J(E)<0)
\end{align}
where $N_{\star,\mathrm{tot}}$ is the total number of star particles belonging to the galaxy, $J_z$ is an individual particle's component of angular momentum about the principle rotation axis (computed from the angular momenta of all stars within 10 half-mass radii), and $J(E)$ is the maximum angular momentum of stellar particles ranked by binding energy ($U_{\mathrm{gravity}} + \nu^2$) within 50 ranks of the particle in question. $N_{\star}(J_z / J(E)<0)$ is an approximation to the number of stars whose motions are not coherent with the bulk rotation. Because the velocities in the bulge are expected to be normally distributed about $J_z / J(E)=0$, symmetry provides that $2\times N_{\star}(J_z / J(E)<0)$ should approximate the number of stars in the bulge of a galaxy -- which can be normalized by the total number of stars for the kinematic bulge-to-total ratio.

\begin{figure*}
	\center\includegraphics[width=0.8\linewidth]{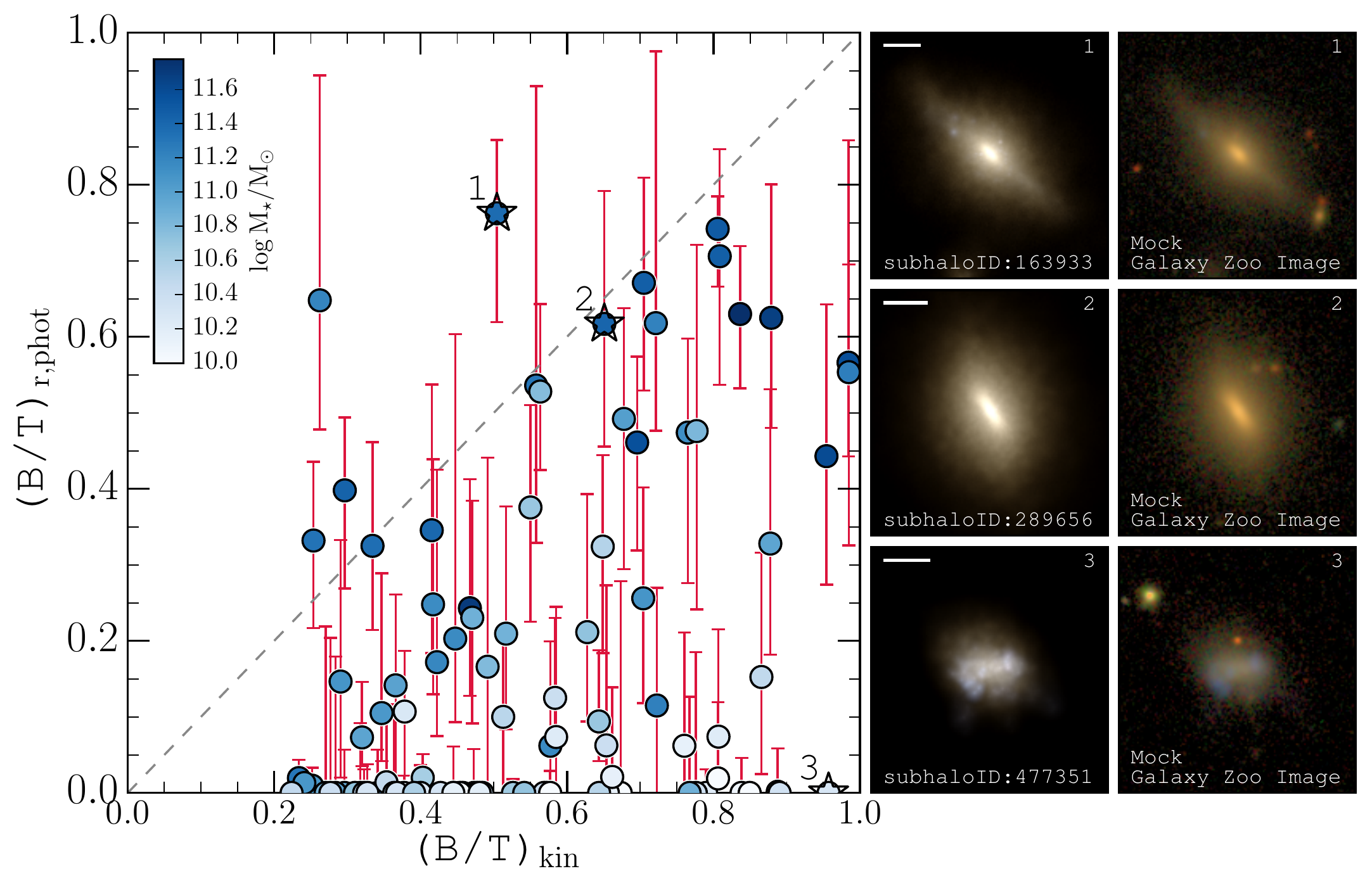}
	\caption[Photometric vs. kinematic bulge-to-total fractions]{Photometric and kinematic bulge-to-total fractions taken from the decompositions results in the \texttt{ASKA} catalog for galaxies in the RIG sample. Points in the left panel show the median photometric $(B/T)_r$ over all placements and camera angles for each galaxy as a function of the kinematic $(B/T)$ derived using Equation \ref{eq:bt_kin} from the orbits of the stellar particles in the simulation. Each point is colour-coded by the total stellar mass of the galaxy, indicated in the colourbar at the top left of the panel. Error bars show the 95\% confidence interval about the medians. The grey dashed line shows the one-to-one relation on this plane. $gri$-colour composites of our synthetic images and Galaxy Zoo \citep{2015MNRAS.454.1886S} images in the right panels correspond to the labeled points (1,2,3) enclosed in star symbols in the left panel. White horizontal lines at the top left of the FoF images measure 10 kpc. The $gri$ false-colour images are shown only for visual impression. Only the $g$ and $r$ band images are used in the modelling.}
    \label{fig:phot_kin}
\end{figure*}

The left panel of Figure \ref{fig:phot_kin} compares the kinematic and photometric estimates of $(B/T)$ using the decompositions from the \texttt{ASKA} catalog (Section \ref{catalogs}). The \texttt{ASKA} galaxies are used to provide a sense of the uncertainties associated with photometric $(B/T)$ by employing the distributions of decomposition results from all placements and camera angles for each galaxy. The vertical position of each point represents the median photometric $(B/T)$ over all placements and camera angles for each galaxy. The error bars show the 95\% range centred on the median of the distribution of estimates. The horizontal position for each system is the kinematic $(B/T)$ derived from the stellar orbits. Note that none of the galaxies in this sample have kinematic bulge fractions less than $(B/T)_{\mathrm{kin}}=0.2$ -- which creates immediate tension with our photometric decomposition results. Many galaxies with high kinematic bulge fractions have no photo-bulge. So whilst the kinematic $(B/T)$ indicate that many galaxies are bulge dominated, the photometric results are completely disk dominated! Furthermore, no discernible correlation is seen between the kinematic and photometric bulge fractions. The results are consistent with previous findings that the photometric bulge-to-total fractions are systematically lower than the kinematic fractions \citep{2010MNRAS.407L..41S,2016MNRAS.459..467O}.

Three RIGs in the left panel of Figure \ref{fig:phot_kin} are highlighted by star symbols and have labels corresponding to image panels to the right. The highlighted RIGs were selected to enable visual inspection of galaxies with $(B/T)_{\mathrm{phot}}>(B/T)_{\mathrm{kin}}$ (upper right row), $(B/T)_{\mathrm{phot}}\approx(B/T)_{\mathrm{kin}}$ (middle right row), $(B/T)_{\mathrm{phot}}<(B/T)_{\mathrm{kin}}$ (bottom right row). The panels show $gri$ composites of our synthetic images and mock SDSS Galaxy Zoo visual classification images\footnote{The mock SDSS Galaxy Zoo images were designed to enable consistent \emph{visual} classifications with observed SDSS galaxies -- so they are Illustris galaxies placed in real image fields, but are not convolved with the SDSS PSF or inserted into a SDSS fields in a way that reproduces crowding, resolution, and sky brightness statistics. These higher-order biases are unimportant for consistency in visual classifications of galaxies (as most galaxies in the vicinity of closely projected sources are rejected from visual classification samples) but are important in decompositions \citep{Bottrell2016}.} with realism from \cite{2015MNRAS.454.1886S} for visual impression of each highlighted RIG. Visual inspection of the morphology of the RIG with $(B/T)_{\mathrm{phot}}>(B/T)_{\mathrm{kin}}$ shows that it has a strong bulge component but that is embedded within a disc (edge-on in this camera angle). The uncertainties from the distribution of decomposition results are not consistent with the kinematic estimate. However, the photometric $(B/T)$ fraction for this galaxy is reconcilable with its visual appearance. The photometric $(B/T)$ fraction for the galaxy with $(B/T)_{\mathrm{phot}}\approx(B/T)_{\mathrm{kin}}$ in the middle row of images in Figure \ref{fig:phot_kin} is also visually reconcilable with the photometry. While the galaxy appears to contain a bar that may affect the photometric (B/T), it is consistent with the kinematically derived quantity. The galaxy shown in the bottom row of images in Figure \ref{fig:phot_kin} is most intriguing. The galaxy shown in the bottom row has $(B/T)_{\mathrm{phot}}=0$ for all placements but $(B/T)_{\mathrm{kin}}\approx1$. However, there is no visual presence of a bulge or a disc -- yet the kinematic information indicates that it is almost a pure bulge. Figure \ref{fig:phot_kin} shows that photometrically derived morphologies can achieve similar results to the kinematics. However, photometric $(B/T)$ estimates for the majority of galaxies in the RIG sample are systematically lower than their kinematic counterparts. 

The markers corresponding to each RIG in Figure \ref{fig:phot_kin} are coloured according to their total stellar masses. Colour-coding of the masses enables inspection of the dependence on stellar mass for the photometric and kinematic bulge-to-total estimates. As expected from Figure \ref{fig:sdss_illustris_bulges}, only galaxies with stellar masses  $\log \mathrm{M}_{\star}/\mathrm{M}_{\odot}\gtrsim11$ contain appreciable photometric bulge fractions. Furthermore, galaxies with low stellar masses $\log \mathrm{M}_{\star}/\mathrm{M}_{\odot}\lesssim10.5$ have the largest systematic errors between the photometric and kinematic estimates for (B/T). The small number of photometric bulges at low masses and the presence of kinematic bulges corroborates with the bias expected from particle resolution. However, the galaxy shown in bottom right row of images in Figure \ref{fig:phot_kin} does not generate confidence in the kinematic estimates for low-mass galaxies. Alternatively, the galaxy in the bottom right row of images in Figure \ref{fig:phot_kin} tells us that kinematics sometimes have nothing to do with visual or photometric morphology! Either way, the high kinematic bulge fractions of galaxies with no visual bulge such as seen in the bottom right row of images in Figure \ref{fig:phot_kin} make pinning particle resolution as the driving bias for reducing photometric $(B/T)$ estimates more challenging. Still, the stronger correlation between kinematic and photometric $(B/T)$ for galaxies with $\log \mathrm{M}_{\star}/\mathrm{M}_{\odot}\gtrsim11$ and the systematically low photometric $(B/T)$ for galaxies with masses $\log \mathrm{M}_{\star}/\mathrm{M}_{\odot}\lesssim10.5$ presents a strong case for the suppression of photometric bulge-to-total fractions by particle mass resolution limitations. 

\begin{figure}
	\center\includegraphics[width=1.04\linewidth]{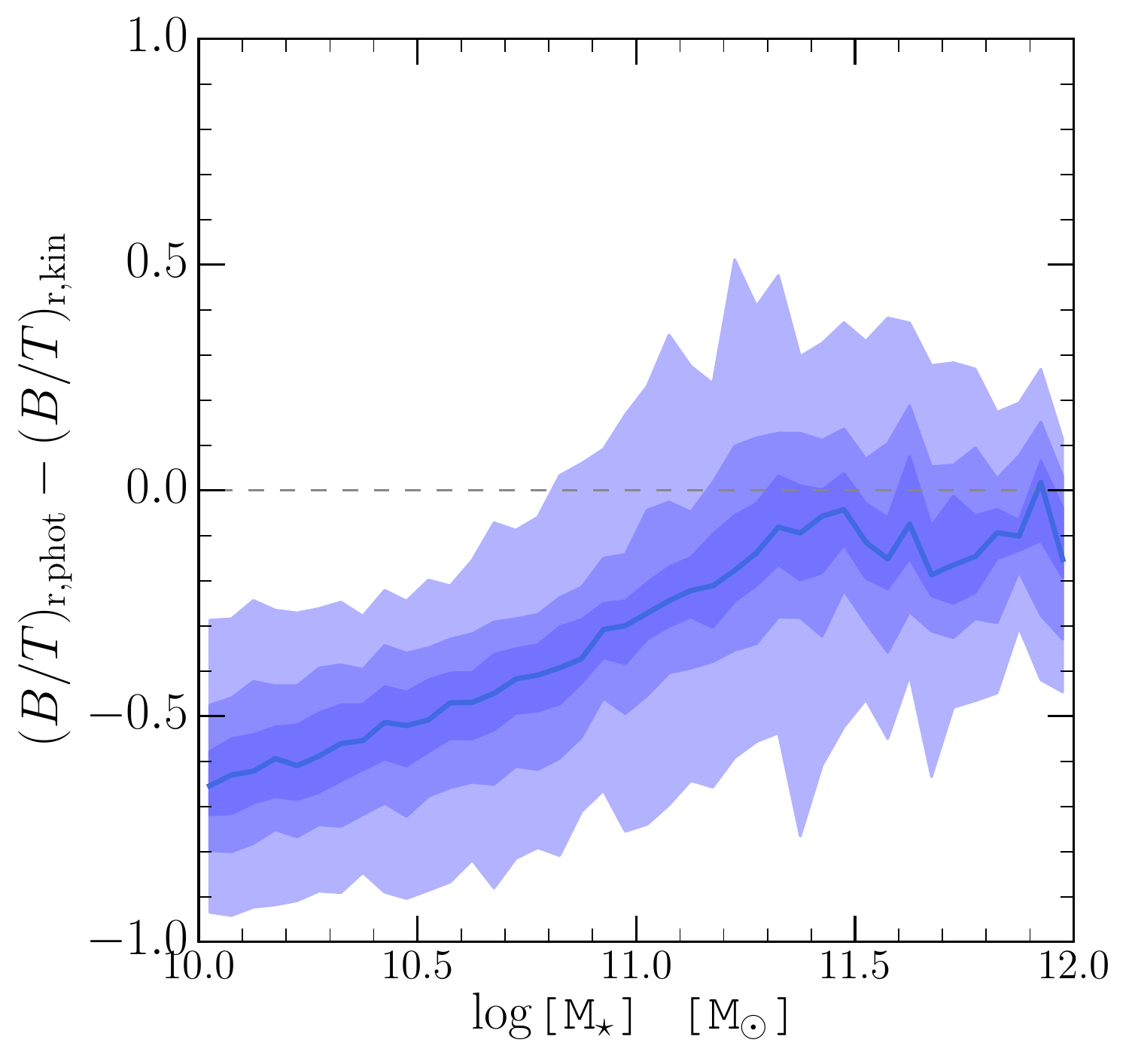}
	\caption[Photometric vs. kinematic bulge-to-total fractions as a function of stellar mass]{The difference in photometric and kinematic bulge fractions, $\Delta(B/T)$, for the full sample of Illustris galaxies shown as a function of total stellar mass. Filled contours show the $67^\mathrm{th}$, $86^\mathrm{th}$, and $95^\mathrm{th}$ percentiles about the median in bins of total stellar mass. The solid dark blue line shows the median. $(B/T)_{r,\mathrm{phot}}$ estimates were taken from the \texttt{DISTINCT} catalog. There is a strong dependence in the correlation between photometric and kinematic $(B/T)$ and total stellar mass. At low masses, $\log \mathrm{M}_{\star}/\mathrm{M}_{\odot}\lesssim11$, photometric $(B/T)$ are systematically smaller. With increasing total stellar mass, the coupling between the photometric and kinematic bulge fraction improves substantially.}
    \label{fig:phot_kin_mass}
\end{figure}

Any relationship between the correlation of kinematic and photometric $(B/T)$ with total stellar mass can be examined using the full Illustris sample (ie. the \texttt{DISTINCT} catalog). Figure \ref{fig:phot_kin_mass} shows the difference between the kinematic and photometric bulge fractions, $\Delta (B/T)$, as a function of total stellar mass for all Illustris galaxies. The suspicion of a mass dependence in $\Delta (B/T)$ from the representative sample in Figure \ref{fig:phot_kin} is confirmed in Figure \ref{fig:phot_kin_mass}. Galaxies at low stellar masses, $\log \mathrm{M}_{\star}/\mathrm{M}_{\odot}\lesssim11$, have systematically lower photometric bulge fractions than inferred from the stellar kinematics. However, increasing total stellar mass yields increasing agreement between the kinematics and photometry. Indeed, at $\log \mathrm{M}_{\star}/\mathrm{M}_{\odot}\gtrsim11.2$, photometric and kinematic $(B/T)$ are broadly consistent.

\section{Discussion}\label{sec:discussion}

\subsection{Remaining challenges to realism: Dust}\label{sec:dust}

Our comparisons of the global size-luminosity relations of SDSS and Illustris indicate that there is no single sufficient explanation for their discrepancy. In Sections \ref{sec:slrs} and \ref{sec:bulgedisc}, we demonstrated that (1) internal segmentation had little overall effect in our comparisons; and (2) the difference in morphologies between the SDSS and Illustris sample has a crucial role in generating the discrepancy. However, Figure \ref{fig:sizemag_btmatch} showed that while matching by morphology improves agreement in the slope and offset of the size-luminosity relations, galaxies in Illustris remain slightly brighter and larger on average for the same stellar masses and bulge fractions. Differences in each may be explained in part by contributions from neglecting treatment of dust in our mock observations of Illustris galaxies. However, the inability to resolve dust physics with the spatial resolution achievable for simulations of the scale of Illustris makes it difficult to examine the role of dust quantitatively. Our comparison of the size-luminosity relations of Illustris and the SDSS galaxies are therefore complicated by the presence of dust in the real universe -- which is not distributed uniformly within galaxies (e.g., see \citealt{2016MNRAS.457.3775M} and references therein).

\cite{2010MNRAS.403.2053G} showed that when dust is present in galaxies, measurements of galaxy properties with bulge+disc decompositions are affected. In their study, disc scale-lengths of analytic bulge+disc systems with dust were systematically over-estimated and this was exacerbated by inclination (with edge-on discs biased most strongly). Meanwhile, the \Sersic{} indices and effective radii of bulges and spheroids were systematically underestimated \citep{2010MNRAS.403.2053G}. In populations of discs and spheroids, the effects of dust may therefore serve to increase the scatter \emph{and} modify the scaling between size and luminosity. However, the differences in size at the low-luminosity end of Figure \ref{fig:sizemag} cannot be reconciled with this dust model because, at low luminosities, galaxies in the SDSS are systematically smaller than in Illustris for the same luminosities -- whereas real galaxies with dust should cause over-estimates of disc sizes. For dust to cause this shift would first require a significantly larger population of low-luminosity spheroids whose sizes would be systematically underestimated due to dust. Therefore, while dust may partially explain the offset in luminosities between Illustris and the SDSS, a difference in morphologies between the two populations is first necessary to cause the changes to the scaling between size and luminosity from dust. Ultimately, the absence of dust in the simulation and radiative transfer code used to produce the synthetic images presents a limitation in the realism of the mock observations. However, the choice to not employ a dust model in the radiative transfer is motivated by the uncertainties involved in the distribution of dust in galaxies. Future analysis of the biases from dust-inclusive radiative transfer for our measurements would yield interesting results that could be compared with the dust-less models, but is beyond the scope of the current work. 

Furthermore, a deficit of photometrically derived bulges relative to observed galaxies is not expected for simulated galaxies in which there is \emph{no dust}. Broadly following the arguments in the previous paragraph, the inclusion of a dust-model in the radiative transfer would serve to further systematically under-estimate $(B/T)$ estimates in Illustris by reducing the overall brightness of bulges and the pixels corresponding to the peak of the bulge surface brightness distribution \citep{2008MNRAS.388.1708G,2010MNRAS.403.2053G}. The de Vaucouleurs $n=4$ model for the bulge is strongly dependent on the surface brightness from the inner 100 pc of a galaxies light profile. Attenuation of the light by dust at the centre of the bulge drives the free \Sersic{} index and bulge half-light radius down in pure \Sersic{} models \citep{2010MNRAS.403.2053G}. In fixed $n$ bulge+disc decompositions the bulge model brightness is forced down to accommodate the decrease in flux from the central pixels and the exponential disc model is driven up. The combined effects lead to a reduction in bulge integrated magnitude and half-light radius, an increase in disc integrated magnitude and half-light radius, and a corresponding reduction in bulge-to-total fraction. The inclusion of dust tends to weaken the photo-bulge relative to the disc. Therefore, the exclusion of dust in the radiative transfer should not cause the photo-bulge deficit.


\subsection{Disconnect between kinematics and photometry}

Future directions in investigating how well the photometric structure of Illustris galaxies actually reproduce the structure of real galaxies will require better treatment of dust radiative transfer, particle resolution and smoothing of stellar light, and comparisons with observations. The facts that (1) high-mass galaxies in Illustris contain photo-bulges that are consistent with the kinematics and (2) mock observations from high- particle resolution simulations also produce bulges \citep{2014MNRAS.440L..51C} support a scenario in which resolution plays a role in interpretation of bulges in mock observations of simulated galaxies. To test the hypothesis that spatial or particle resolution is preventing adequate sampling of the bulge component, a deeper examination of alternative SLD schemes and, more directly, comparisons with zoom-in simulations of Illustris are required. In particular, a recent zoom-in simulation of Illustris using the same model and hydrodynamics improved the particle resolution by 40$\times$ the resolution of the full volume \citep{2016arXiv160408205S}. The zoom-in would place the particle resolution of the zoomed-in low mass galaxies, $\log \mathrm{M}_{\star}/\mathrm{M}_{\odot} \lesssim 10.5$, on level-ground with galaxies at high mass in our current comparison -- which appear to contain more substantial numbers of galaxies with bulges. If bulges can be resolved in synthetic images of the zoomed-in low-mass galaxies, then new constraints can be placed on the necessary resolution to resolve bulges for realistic comparisons with observations. 

Furthermore, the nature of whether photo-bulges genuinely correlate with the kinematic bulges is another test that may be performed with a comparison of both kinematic and photometric structural estimates of individual galaxies in the zoom-in and the full volume. Given that our lowest mass galaxies had the largest discrepancies in the kinematic and photometric bulge fractions, one could tackle the question of whether an increase by $400\%$ in particle resolution can alleviate the discrepancy. 

It may also be possible to explore alternative kinematic definitions of a bulge. In this paper, we used a simple prescription that assumes that the angular momentum distribution of the bulge component is symmetric about the principle angular momentum axis of the galaxy -- centred at the minimum in the gravitational potential. In this way, bulge fraction can be easily computed assuming this symmetry, since stellar particles belonging to a rotationally supported disc should have an angular momentum distribution that should be offset from zero. However, there may be several caveats to this definition. A comprehensive study of caveats to this definition is beyond the scope of this paper. Still, one of them could be the stellar clumps in Illustris galaxies identified in \cite{Bottrell2016}. Should a handful of large stellar clumps have orbits in which a large amount of the angular momentum is radial, this would cause the kinematic bulge fraction to systematically increase -- despite the clumps having no physical connection to a stellar bulge. The contributions of these clumps to the angular momentum distributions of higher mass galaxies may be smaller, leading to generally better relationships between kinematic and photometric bulges. Although speculative at this point, now that we have seen that coupling between the kinematic and photometric bulge fractions is possible, it may be a good opportunity to review and improve upon conventional definitions of the bulge inferred from simulations.

\section{summary}\label{sec:summary}

In this second paper in a series, we have employed our new procedure described in \cite{Bottrell2016} for deriving image-based quantitative morphologies of simulated galaxies in a fair comparison with observations. In this section, the results of a comparison between properties derived from bulge+disc decompositions of galaxies from the Illustris simulation and the SDSS are summarized.

\subsection{Comparison with SDSS galaxies}

The bulge+disc decomposition results from the \texttt{DISTINCT} catalog \citep{Bottrell2016} were used to perform a comparison with the observed size-luminosity and (B/T)-stellar mass relations. Comparisons between simulated and real galaxy size-luminosity relations have been previously performed (e.g., \citealt{2011ApJ...728...51B,2014MNRAS.440L..51C,2015MNRAS.454.1886S,2015arXiv151005645F}) but may be compromised by at least one of the following factors: (1) the limitations on statistical meaningfulness in comparisons to observations due to small simulated galaxy samples; (2) inconsistent derivations of simulated and observed galaxy properties; (3) incomplete observational realism that biases the distributions of derived properties of simulated galaxies in comparisons with observations. Each of these caveats is addressed by using mock observations of a representative population of simulated galaxies, applying extensive observational realism to enable an unbiased image-based comparison, and employing identical methods for deriving galaxy properties in simulated and observed galaxies.

\begin{itemize}

\item \emph{Size-luminosity relation}: Illustris galaxies were matched by stellar mass to the stellar mass distribution of the SDSS -- taking the SDSS galaxy population over $0.04<z<0.06$ with a lower mass cut of $\log \mathrm{M}_{\star}/\mathrm{M}_{\odot} > 10$ as the comparison sample. In Section \ref{sec:slrs}, we compared the size-luminosity relations of the SDSS and Illustris for the matched sample and showed that Illustris galaxies are generally larger and brighter for the same stellar masses as galaxies from the SDSS. Furthermore, the correlation between size and luminosity is not as strong in Illustris (and appears flat) relative to the SDSS relation. We concluded that such a discrepancy could not be explained by the known biases from internal segmentation of the galaxy surface brightness distributions identified in \cite{Bottrell2016}. In Section \ref{sec:bulgedisc}, the morphological dependence of the size-luminosity relations in Illustris and SDSS was examined using our bulge-to-total fractions. While Illustris qualitatively reproduces the observed trend of increasing slope and decreasing normalization with increasing bulge fractions, the size-luminosity relations of Illustris are quantitatively distinct, having smaller slopes and higher normalizations across all $(B/T)$ classifications (Figure \ref{fig:sizemag_bt}).

\item \emph{Bulge and disc morphologies}: Distributions of $(B/T)$ as a function of total stellar mass were compared using the mass-matched samples from the size-luminosity comparison. We showed that Illustris is dominated by disc-dominated morphologies at all masses in the sample -- whereas the SDSS demonstrates diverse morphologies. Still, Illustris contains bulges-dominated galaxies, but the relationship between stellar mass and $(B/T)$ is stronger than in the observations (i.e. Illustris contains too many discs at low mass and only high-mass galaxies contain appreciable bulge fractions). The size-luminosity relations of bulge- and disc- dominated galaxies differ significantly in observations. These results hinted that the morphological differences between Illustris and SDSS samples were affecting the comparison of their size-luminosity relations.

\item \emph{Size-luminosity relation -- Impact of morphology}: The size-luminosity relations of SDSS and Illustris were revisited -- this time matching by stellar mass \emph{and} $(B/T)$ morphology by re-sampling SDSS galaxies to match the (B/T)-stellar mass distribution of Illustris. The comparison demonstrated that, indeed, the discrepancy in the previous comparison of the size-luminosity relations owed in large part to the fact that Illustris contained predominantly disc-dominated galaxies in that sample. By additionally matching by morphology, the agreement between the size-luminosity relations (which is essentially the disc size-luminosity relation) is significantly improved -- leaving a reduced magnitude and size offset between the relations. The remaining offset is difficult to characterize without a detailed quantification of the effects of dust in the creation of the synthetic images and on our decomposition results.

\end{itemize}

\subsection{The deficit of bulge-dominated galaxies in Illustris at low stellar mass}

Based on our decompositions with \textsc{gim2d}, Illustris contains too few bulge/spheroid-dominated galaxies at low stellar masses, $\log \mathrm{M}_{\star}/\mathrm{M}_{\odot} \lesssim 11$, relative to the observations and only has appreciable populations of galaxies with bulges at $\log \mathrm{M}_{\star}/\mathrm{M}_{\odot} \gtrsim 11$. The deficit of bulge-dominated galaxies contrasts with previous generations of simulations that tended to produce galaxies that were too compact, dense, and rotated too quickly (e.g., \citealt{1999ApJ...513..555S,2000ApJ...538..477N,2001ApJ...554..114E,2003ApJ...591..499A,2004ApJ...607..688G,piontek2009angular,2012MNRAS.423.1726S}. A relationship between bulge fraction and stellar mass is not surprising in a framework of galaxy evolution that is based on hierarchical assembly of galaxies through mergers. However, the significantly stronger dependence on bulge-fraction with total stellar mass in Illustris, relative to the observations, is puzzling. Still, several scenarios may provide some explanation. First, the ability to resolve the bulge for galaxies with the spatial resolution that is limited by the number of stellar particles and/or method for distributing stellar light in galaxies with low total stellar mass. And second, the adequacy of the mechanisms by which bulges form and survive within the simulation model. Application of the methods utilized in this paper to zoom-in cosmological simulations with the same physical model at high-resolution (e.g., \citealt{2016arXiv160408205S}) and to alternative models (e.g., EAGLE: \citealt{2015MNRAS.446..521S}) may yield insight into the validity of these hypotheses.

Lastly, the photometric bulge-to-total fractions of Illustris galaxies were compared with the bulge fractions derived from the internal kinematics of simulated galaxies. Confirming previous work using a larger sample and similar resolution \citep{2010MNRAS.407L..41S,2016MNRAS.459..467O}, we showed in Section \ref{phot_vs_kin} that the photometric estimates for $(B/T)$ are systematically lower than the kinematic estimates. In our first look using a representative sample of Illustris galaxies, no discernible correlation between the photometric and kinematic $(B/T)$ is seen. However, taking all galaxies in Illustris, we showed that while galaxies with $\log \mathrm{M}_{\star}/\mathrm{M}_{\odot} \lesssim 11$ have photometric $(B/T)$ that are systematically lower than the kinematic $(B/T)$, galaxies with higher stellar masses demonstrated broad consistency between photometric and kinematic bulge fractions. We showed that there is a strong relationship between the correlation of photometric and kinematic $(B/T)$ and total stellar mass -- with the correlation improving with increasing masses.
Several low-mass galaxies $\log \mathrm{M}_{\star}/\mathrm{M}_{\odot} \lesssim 10.5$ with high kinematic $(B/T)$ and no visible photo-bulge were inspected -- implying that (a) the spatial resolution is insufficient in these galaxies to resolve the bulge; (b) the kinematic estimate for the bulge that we employed does not always reflect the true presence of a bulge; (c) there is no underlying connection between kinematics and visual or photometric morphology. A combination of (a) and (b) is also possible. In such a scenario, galaxies that are poorly resolved (both spatially in the images and by particles in the kinematics) may have reduced photometric estimates of $(B/T)$ \emph{and} have intrinsically large uncertainties in the kinematic estimates.

\section*{Acknowledgements}

We thank the reviewer for suggestions which greatly improved the quality of this paper. We thank Greg Snyder for useful discussions and input. PT acknowledges support for Program number HST-HF2-51384.001-A was provided by NASA through a grant from the Space Telescope Science Institute, which is operated by the Association of Universities for Research in Astronomy, Incorporated, under NASA contract NAS5-26555. This research made use of a University of Victoria computing facility funded by grants from the Canadian Foundation for Innovation and the British Columbia Knowledge and Development Fund. We thank the system administrators of this facility for their gracious support. Funding for the Sloan Digital Sky Survey IV has been provided by
the Alfred P. Sloan Foundation, the U.S. Department of Energy Office of
Science, and the Participating Institutions. SDSS-IV acknowledges
support and resources from the Center for High-Performance Computing at
the University of Utah. The SDSS web site is www.sdss.org. SDSS-IV is managed by the Astrophysical Research Consortium for the 
Participating Institutions of the SDSS Collaboration including the 
Brazilian Participation Group, the Carnegie Institution for Science, 
Carnegie Mellon University, the Chilean Participation Group, the French Participation Group, Harvard-Smithsonian Center for Astrophysics, 
Instituto de Astrof\'isica de Canarias, The Johns Hopkins University, 
Kavli Institute for the Physics and Mathematics of the Universe (IPMU) / 
University of Tokyo, Lawrence Berkeley National Laboratory, 
Leibniz Institut f\"ur Astrophysik Potsdam (AIP),  
Max-Planck-Institut f\"ur Astronomie (MPIA Heidelberg), 
Max-Planck-Institut f\"ur Astrophysik (MPA Garching), 
Max-Planck-Institut f\"ur Extraterrestrische Physik (MPE), 
National Astronomical Observatory of China, New Mexico State University, 
New York University, University of Notre Dame, 
Observat\'ario Nacional / MCTI, The Ohio State University, 
Pennsylvania State University, Shanghai Astronomical Observatory, 
United Kingdom Participation Group,
Universidad Nacional Aut\'onoma de M\'exico, University of Arizona, 
University of Colorado Boulder, University of Oxford, University of Portsmouth, 
University of Utah, University of Virginia, University of Washington, University of Wisconsin, 
Vanderbilt University, and Yale University.

\bibliographystyle{mnras}
\bibliography{Bib/references_full.bib} 

\begin{thebibliography}{}
\makeatletter
\relax
\def\mn@urlcharsother{\let\do\@makeother \do\$\do\&\do\#\do\^\do\_\do\%\do\~}
\def\mn@doi{\begingroup\mn@urlcharsother \@ifnextchar [ {\mn@doi@}
  {\mn@doi@[]}}
\def\mn@doi@[#1]#2{\def\@tempa{#1}\ifx\@tempa\@empty \href
  {http://dx.doi.org/#2} {doi:#2}\else \href {http://dx.doi.org/#2} {#1}\fi
  \endgroup}
\def\mn@eprint#1#2{\mn@eprint@#1:#2::\@nil}
\def\mn@eprint@arXiv#1{\href {http://arxiv.org/abs/#1} {{\tt arXiv:#1}}}
\def\mn@eprint@dblp#1{\href {http://dblp.uni-trier.de/rec/bibtex/#1.xml}
  {dblp:#1}}
\def\mn@eprint@#1:#2:#3:#4\@nil{\def\@tempa {#1}\def\@tempb {#2}\def\@tempc
  {#3}\ifx \@tempc \@empty \let \@tempc \@tempb \let \@tempb \@tempa \fi \ifx
  \@tempb \@empty \def\@tempb {arXiv}\fi \@ifundefined
  {mn@eprint@\@tempb}{\@tempb:\@tempc}{\expandafter \expandafter \csname
  mn@eprint@\@tempb\endcsname \expandafter{\@tempc}}}

\bibitem[\protect\citeauthoryear{{Abadi}, {Navarro}, {Steinmetz}  \&
  {Eke}}{{Abadi} et~al.}{2003}]{2003ApJ...591..499A}
{Abadi} M.~G.,  {Navarro} J.~F.,  {Steinmetz} M.,   {Eke} V.~R.,  2003, \mn@doi
  [\apj] {10.1086/375512}, \href
  {http://adsabs.harvard.edu/abs/2003ApJ...591..499A} {591, 499}

\bibitem[\protect\citeauthoryear{{Agertz}, {Teyssier}  \& {Moore}}{{Agertz}
  et~al.}{2011}]{2011MNRAS.410.1391A}
{Agertz} O.,  {Teyssier} R.,   {Moore} B.,  2011, \mn@doi [\mnras]
  {10.1111/j.1365-2966.2010.17530.x}, \href
  {http://adsabs.harvard.edu/abs/2011MNRAS.410.1391A} {410, 1391}

\bibitem[\protect\citeauthoryear{{Aumer}, {White}, {Naab}  \&
  {Scannapieco}}{{Aumer} et~al.}{2013}]{2013MNRAS.434.3142A}
{Aumer} M.,  {White} S.~D.~M.,  {Naab} T.,   {Scannapieco} C.,  2013, \mn@doi
  [\mnras] {10.1093/mnras/stt1230}, \href
  {http://adsabs.harvard.edu/abs/2013MNRAS.434.3142A} {434, 3142}

\bibitem[\protect\citeauthoryear{{Baldry} et~al.,}{{Baldry}
  et~al.}{2012}]{2012MNRAS.421..621B}
{Baldry} I.~K.,  et~al., 2012, \mn@doi [\mnras]
  {10.1111/j.1365-2966.2012.20340.x}, \href
  {http://adsabs.harvard.edu/abs/2012MNRAS.421..621B} {421, 621}

\bibitem[\protect\citeauthoryear{{Behroozi}, {Wechsler}  \&
  {Conroy}}{{Behroozi} et~al.}{2013}]{2013ApJ...770...57B}
{Behroozi} P.~S.,  {Wechsler} R.~H.,   {Conroy} C.,  2013, \mn@doi [\apj]
  {10.1088/0004-637X/770/1/57}, \href
  {http://adsabs.harvard.edu/abs/2013ApJ...770...57B} {770, 57}

\bibitem[\protect\citeauthoryear{{Bernardi}, {Meert}, {Sheth}, {Vikram},
  {Huertas-Company}, {Mei}  \& {Shankar}}{{Bernardi}
  et~al.}{2013}]{2013MNRAS.436..697B}
{Bernardi} M.,  {Meert} A.,  {Sheth} R.~K.,  {Vikram} V.,  {Huertas-Company}
  M.,  {Mei} S.,   {Shankar} F.,  2013, \mn@doi [\mnras]
  {10.1093/mnras/stt1607}, \href
  {http://adsabs.harvard.edu/abs/2013MNRAS.436..697B} {436, 697}

\bibitem[\protect\citeauthoryear{{Blanton} \& {Moustakas}}{{Blanton} \&
  {Moustakas}}{2009}]{2009ARA&A..47..159B}
{Blanton} M.~R.,  {Moustakas} J.,  2009, \mn@doi [\araa]
  {10.1146/annurev-astro-082708-101734}, \href
  {http://adsabs.harvard.edu/abs/2009ARA%26A..47..159B} {47, 159}

\bibitem[\protect\citeauthoryear{{Bluck} et~al.,}{{Bluck}
  et~al.}{2016}]{2016MNRAS.462.2559B}
{Bluck} A.~F.~L.,  et~al., 2016, \mn@doi [\mnras] {10.1093/mnras/stw1665},
  \href {http://adsabs.harvard.edu/abs/2016MNRAS.462.2559B} {462, 2559}

\bibitem[\protect\citeauthoryear{{Bottrell}, {Torrey}, {Simard}  \&
  {Ellison}}{{Bottrell} et~al.}{2017}]{Bottrell2016}
{Bottrell} C.,  {Torrey} P.,  {Simard} L.,   {Ellison} S.~L.,  2017, submitted
  to \mnras

\bibitem[\protect\citeauthoryear{{Brook} et~al.,}{{Brook}
  et~al.}{2011}]{2011MNRAS.415.1051B}
{Brook} C.~B.,  et~al., 2011, \mn@doi [\mnras]
  {10.1111/j.1365-2966.2011.18545.x}, \href
  {http://adsabs.harvard.edu/abs/2011MNRAS.415.1051B} {415, 1051}

\bibitem[\protect\citeauthoryear{{Brook}, {Stinson}, {Gibson}, {Ro{\v s}kar},
  {Wadsley}  \& {Quinn}}{{Brook} et~al.}{2012a}]{2012MNRAS.419..771B}
{Brook} C.~B.,  {Stinson} G.,  {Gibson} B.~K.,  {Ro{\v s}kar} R.,  {Wadsley}
  J.,   {Quinn} T.,  2012a, \mn@doi [\mnras]
  {10.1111/j.1365-2966.2011.19740.x}, \href
  {http://adsabs.harvard.edu/abs/2012MNRAS.419..771B} {419, 771}

\bibitem[\protect\citeauthoryear{{Brook}, {Stinson}, {Gibson}, {Wadsley}  \&
  {Quinn}}{{Brook} et~al.}{2012b}]{2012MNRAS.424.1275B}
{Brook} C.~B.,  {Stinson} G.,  {Gibson} B.~K.,  {Wadsley} J.,   {Quinn} T.,
  2012b, \mn@doi [\mnras] {10.1111/j.1365-2966.2012.21306.x}, \href
  {http://adsabs.harvard.edu/abs/2012MNRAS.424.1275B} {424, 1275}

\bibitem[\protect\citeauthoryear{{Brooks}}{{Brooks}}{2010}]{2010ASPC..432...17B}
{Brooks} A.,  2010, in {Stanford} L.~M.,  {Green} J.~D.,  {Hao} L.,   {Mao} Y.,
   eds,  Astronomical Society of the Pacific Conference Series Vol. 432, New
  Horizons in Astronomy: Frank N. Bash Symposium 2009. p.~17 (\mn@eprint
  {arXiv} {1003.3882})

\bibitem[\protect\citeauthoryear{{Brooks} et~al.,}{{Brooks}
  et~al.}{2011}]{2011ApJ...728...51B}
{Brooks} A.~M.,  et~al., 2011, \mn@doi [\apj] {10.1088/0004-637X/728/1/51},
  \href {http://adsabs.harvard.edu/abs/2011ApJ...728...51B} {728, 51}

\bibitem[\protect\citeauthoryear{{Chabrier}}{{Chabrier}}{2003}]{2003PASP..115..763C}
{Chabrier} G.,  2003, \mn@doi [\pasp] {10.1086/376392}, \href
  {http://adsabs.harvard.edu/abs/2003PASP..115..763C} {115, 763}

\bibitem[\protect\citeauthoryear{{Christensen}, {Brooks}, {Fisher},
  {Governato}, {McCleary}, {Quinn}, {Shen}  \& {Wadsley}}{{Christensen}
  et~al.}{2014}]{2014MNRAS.440L..51C}
{Christensen} C.~R.,  {Brooks} A.~M.,  {Fisher} D.~B.,  {Governato} F.,
  {McCleary} J.,  {Quinn} T.~R.,  {Shen} S.,   {Wadsley} J.,  2014, \mn@doi
  [\mnras] {10.1093/mnrasl/slu020}, \href
  {http://adsabs.harvard.edu/abs/2014MNRAS.440L..51C} {440, L51}

\bibitem[\protect\citeauthoryear{{Conselice}}{{Conselice}}{2006}]{2006MNRAS.373.1389C}
{Conselice} C.~J.,  2006, \mn@doi [\mnras] {10.1111/j.1365-2966.2006.11114.x},
  \href {http://adsabs.harvard.edu/abs/2006MNRAS.373.1389C} {373, 1389}

\bibitem[\protect\citeauthoryear{{Courteau}, {Dutton}, {van den Bosch},
  {MacArthur}, {Dekel}, {McIntosh}  \& {Dale}}{{Courteau}
  et~al.}{2007}]{2007ApJ...671..203C}
{Courteau} S.,  {Dutton} A.~A.,  {van den Bosch} F.~C.,  {MacArthur} L.~A.,
  {Dekel} A.,  {McIntosh} D.~H.,   {Dale} D.~A.,  2007, \mn@doi [\apj]
  {10.1086/522193}, \href {http://adsabs.harvard.edu/abs/2007ApJ...671..203C}
  {671, 203}

\bibitem[\protect\citeauthoryear{{Crain} et~al.,}{{Crain}
  et~al.}{2009}]{2009MNRAS.399.1773C}
{Crain} R.~A.,  et~al., 2009, \mn@doi [\mnras]
  {10.1111/j.1365-2966.2009.15402.x}, \href
  {http://adsabs.harvard.edu/abs/2009MNRAS.399.1773C} {399, 1773}

\bibitem[\protect\citeauthoryear{{Crain} et~al.,}{{Crain}
  et~al.}{2015}]{2015MNRAS.450.1937C}
{Crain} R.~A.,  et~al., 2015, \mn@doi [\mnras] {10.1093/mnras/stv725}, \href
  {http://adsabs.harvard.edu/abs/2015MNRAS.450.1937C} {450, 1937}

\bibitem[\protect\citeauthoryear{{Davis}, {Efstathiou}, {Frenk}  \&
  {White}}{{Davis} et~al.}{1985}]{1985ApJ...292..371D}
{Davis} M.,  {Efstathiou} G.,  {Frenk} C.~S.,   {White} S.~D.~M.,  1985,
  \mn@doi [\apj] {10.1086/163168}, \href
  {http://adsabs.harvard.edu/abs/1985ApJ...292..371D} {292, 371}

\bibitem[\protect\citeauthoryear{{Dopita} et~al.,}{{Dopita}
  et~al.}{2005}]{2005ApJ...619..755D}
{Dopita} M.~A.,  et~al., 2005, \mn@doi [\apj] {10.1086/423948}, \href
  {http://adsabs.harvard.edu/abs/2005ApJ...619..755D} {619, 755}

\bibitem[\protect\citeauthoryear{{Dopita} et~al.,}{{Dopita}
  et~al.}{2006a}]{2006ApJS..167..177D}
{Dopita} M.~A.,  et~al., 2006a, \mn@doi [\apjs] {10.1086/508261}, \href
  {http://adsabs.harvard.edu/abs/2006ApJS..167..177D} {167, 177}

\bibitem[\protect\citeauthoryear{{Dopita} et~al.,}{{Dopita}
  et~al.}{2006b}]{2006ApJ...647..244D}
{Dopita} M.~A.,  et~al., 2006b, \mn@doi [\apj] {10.1086/505418}, \href
  {http://adsabs.harvard.edu/abs/2006ApJ...647..244D} {647, 244}

\bibitem[\protect\citeauthoryear{{Dutton} et~al.,}{{Dutton}
  et~al.}{2011}]{2011MNRAS.410.1660D}
{Dutton} A.~A.,  et~al., 2011, \mn@doi [\mnras]
  {10.1111/j.1365-2966.2010.17555.x}, \href
  {http://adsabs.harvard.edu/abs/2011MNRAS.410.1660D} {410, 1660}

\bibitem[\protect\citeauthoryear{{Eke}, {Navarro}  \& {Steinmetz}}{{Eke}
  et~al.}{2001}]{2001ApJ...554..114E}
{Eke} V.~R.,  {Navarro} J.~F.,   {Steinmetz} M.,  2001, \mn@doi [\apj]
  {10.1086/321345}, \href {http://adsabs.harvard.edu/abs/2001ApJ...554..114E}
  {554, 114}

\bibitem[\protect\citeauthoryear{{Ellis} et~al.,}{{Ellis}
  et~al.}{2013}]{2013ApJ...763L...7E}
{Ellis} R.~S.,  et~al., 2013, \mn@doi [\apjl] {10.1088/2041-8205/763/1/L7},
  \href {http://adsabs.harvard.edu/abs/2013ApJ...763L...7E} {763, L7}

\bibitem[\protect\citeauthoryear{{Faucher-Gigu{\`e}re}, {Prochaska}, {Lidz},
  {Hernquist}  \& {Zaldarriaga}}{{Faucher-Gigu{\`e}re}
  et~al.}{2008}]{2008ApJ...681..831F}
{Faucher-Gigu{\`e}re} C.-A.,  {Prochaska} J.~X.,  {Lidz} A.,  {Hernquist} L.,
  {Zaldarriaga} M.,  2008, \mn@doi [\apj] {10.1086/588648}, \href
  {http://adsabs.harvard.edu/abs/2008ApJ...681..831F} {681, 831}

\bibitem[\protect\citeauthoryear{{Faucher-Gigu{\`e}re}, {Lidz}, {Zaldarriaga}
  \& {Hernquist}}{{Faucher-Gigu{\`e}re} et~al.}{2009}]{2009ApJ...703.1416F}
{Faucher-Gigu{\`e}re} C.-A.,  {Lidz} A.,  {Zaldarriaga} M.,   {Hernquist} L.,
  2009, \mn@doi [\apj] {10.1088/0004-637X/703/2/1416}, \href
  {http://adsabs.harvard.edu/abs/2009ApJ...703.1416F} {703, 1416}

\bibitem[\protect\citeauthoryear{{Fisher} \& {Drory}}{{Fisher} \&
  {Drory}}{2008}]{2008AJ....136..773F}
{Fisher} D.~B.,  {Drory} N.,  2008, \mn@doi [\aj]
  {10.1088/0004-6256/136/2/773}, \href
  {http://adsabs.harvard.edu/abs/2008AJ....136..773F} {136, 773}

\bibitem[\protect\citeauthoryear{{Fisher} \& {Drory}}{{Fisher} \&
  {Drory}}{2010}]{2010ApJ...716..942F}
{Fisher} D.~B.,  {Drory} N.,  2010, \mn@doi [\apj]
  {10.1088/0004-637X/716/2/942}, \href
  {http://adsabs.harvard.edu/abs/2010ApJ...716..942F} {716, 942}

\bibitem[\protect\citeauthoryear{{Fisher}, {Bolatto}, {Drory}, {Combes},
  {Blitz}  \& {Wong}}{{Fisher} et~al.}{2013}]{2013ApJ...764..174F}
{Fisher} D.~B.,  {Bolatto} A.,  {Drory} N.,  {Combes} F.,  {Blitz} L.,   {Wong}
  T.,  2013, \mn@doi [\apj] {10.1088/0004-637X/764/2/174}, \href
  {http://adsabs.harvard.edu/abs/2013ApJ...764..174F} {764, 174}

\bibitem[\protect\citeauthoryear{{Furlong} et~al.,}{{Furlong}
  et~al.}{2015}]{2015arXiv151005645F}
{Furlong} M.,  et~al., 2015, preprint, \href
  {http://adsabs.harvard.edu/abs/2015arXiv151005645F} {} (\mn@eprint {arXiv}
  {1510.05645})

\bibitem[\protect\citeauthoryear{{Gadotti}}{{Gadotti}}{2009}]{2009MNRAS.393.1531G}
{Gadotti} D.~A.,  2009, \mn@doi [\mnras] {10.1111/j.1365-2966.2008.14257.x},
  \href {http://adsabs.harvard.edu/abs/2009MNRAS.393.1531G} {393, 1531}

\bibitem[\protect\citeauthoryear{{Gadotti}, {Baes}  \& {Falony}}{{Gadotti}
  et~al.}{2010}]{2010MNRAS.403.2053G}
{Gadotti} D.~A.,  {Baes} M.,   {Falony} S.,  2010, \mn@doi [\mnras]
  {10.1111/j.1365-2966.2010.16243.x}, \href
  {http://adsabs.harvard.edu/abs/2010MNRAS.403.2053G} {403, 2053}

\bibitem[\protect\citeauthoryear{{Genel} et~al.,}{{Genel}
  et~al.}{2014}]{2014MNRAS.445..175G}
{Genel} S.,  et~al., 2014, \mn@doi [\mnras] {10.1093/mnras/stu1654}, \href
  {http://adsabs.harvard.edu/abs/2014MNRAS.445..175G} {445, 175}

\bibitem[\protect\citeauthoryear{{Genel}, {Fall}, {Hernquist}, {Vogelsberger},
  {Snyder}, {Rodriguez-Gomez}, {Sijacki}  \& {Springel}}{{Genel}
  et~al.}{2015}]{2015ApJ...804L..40G}
{Genel} S.,  {Fall} S.~M.,  {Hernquist} L.,  {Vogelsberger} M.,  {Snyder}
  G.~F.,  {Rodriguez-Gomez} V.,  {Sijacki} D.,   {Springel} V.,  2015, \mn@doi
  [\apjl] {10.1088/2041-8205/804/2/L40}, \href
  {http://adsabs.harvard.edu/abs/2015ApJ...804L..40G} {804, L40}

\bibitem[\protect\citeauthoryear{{Governato} et~al.,}{{Governato}
  et~al.}{2004}]{2004ApJ...607..688G}
{Governato} F.,  et~al., 2004, \mn@doi [\apj] {10.1086/383516}, \href
  {http://adsabs.harvard.edu/abs/2004ApJ...607..688G} {607, 688}

\bibitem[\protect\citeauthoryear{{Graham}}{{Graham}}{2013}]{2013pss6.book...91G}
{Graham} A.~W.,  2013, {Planets, Stars and Stellar Systems. Volume 6:
  Extragalactic Astronomy and Cosmology}.
Springer, p.~91, \mn@doi{10.1007/978-94-007-5609-0_2}

\bibitem[\protect\citeauthoryear{{Graham} \& {Worley}}{{Graham} \&
  {Worley}}{2008}]{2008MNRAS.388.1708G}
{Graham} A.~W.,  {Worley} C.~C.,  2008, \mn@doi [\mnras]
  {10.1111/j.1365-2966.2008.13506.x}, \href
  {http://adsabs.harvard.edu/abs/2008MNRAS.388.1708G} {388, 1708}

\bibitem[\protect\citeauthoryear{{Groves}, {Dopita}, {Sutherland}, {Kewley},
  {Fischera}, {Leitherer}, {Brandl}  \& {van Breugel}}{{Groves}
  et~al.}{2008}]{2008ApJS..176..438G}
{Groves} B.,  {Dopita} M.~A.,  {Sutherland} R.~S.,  {Kewley} L.~J.,  {Fischera}
  J.,  {Leitherer} C.,  {Brandl} B.,   {van Breugel} W.,  2008, \mn@doi [\apjs]
  {10.1086/528711}, \href {http://adsabs.harvard.edu/abs/2008ApJS..176..438G}
  {176, 438}

\bibitem[\protect\citeauthoryear{{Hayward} \& {Smith}}{{Hayward} \&
  {Smith}}{2015}]{2015MNRAS.446.1512H}
{Hayward} C.~C.,  {Smith} D.~J.~B.,  2015, \mn@doi [\mnras]
  {10.1093/mnras/stu2195}, \href
  {http://adsabs.harvard.edu/abs/2015MNRAS.446.1512H} {446, 1512}

\bibitem[\protect\citeauthoryear{{Hayward}, {Torrey}, {Springel}, {Hernquist}
  \& {Vogelsberger}}{{Hayward} et~al.}{2014}]{2014MNRAS.442.1992H}
{Hayward} C.~C.,  {Torrey} P.,  {Springel} V.,  {Hernquist} L.,
  {Vogelsberger} M.,  2014, \mn@doi [\mnras] {10.1093/mnras/stu957}, \href
  {http://adsabs.harvard.edu/abs/2014MNRAS.442.1992H} {442, 1992}

\bibitem[\protect\citeauthoryear{{Hinshaw} et~al.,}{{Hinshaw}
  et~al.}{2013}]{2013ApJS..208...19H}
{Hinshaw} G.,  et~al., 2013, \mn@doi [\apjs] {10.1088/0067-0049/208/2/19},
  \href {http://adsabs.harvard.edu/abs/2013ApJS..208...19H} {208, 19}

\bibitem[\protect\citeauthoryear{{Hopkins}, {Kere{\v s}}, {O{\~n}orbe},
  {Faucher-Gigu{\`e}re}, {Quataert}, {Murray}  \& {Bullock}}{{Hopkins}
  et~al.}{2014}]{2014MNRAS.445..581H}
{Hopkins} P.~F.,  {Kere{\v s}} D.,  {O{\~n}orbe} J.,  {Faucher-Gigu{\`e}re}
  C.-A.,  {Quataert} E.,  {Murray} N.,   {Bullock} J.~S.,  2014, \mn@doi
  [\mnras] {10.1093/mnras/stu1738}, \href
  {http://adsabs.harvard.edu/abs/2014MNRAS.445..581H} {445, 581}

\bibitem[\protect\citeauthoryear{{Jonsson}}{{Jonsson}}{2006}]{2006MNRAS.372....2J}
{Jonsson} P.,  2006, \mn@doi [\mnras] {10.1111/j.1365-2966.2006.10884.x}, \href
  {http://adsabs.harvard.edu/abs/2006MNRAS.372....2J} {372, 2}

\bibitem[\protect\citeauthoryear{{Jonsson}, {Groves}  \& {Cox}}{{Jonsson}
  et~al.}{2010}]{2010MNRAS.403...17J}
{Jonsson} P.,  {Groves} B.~A.,   {Cox} T.~J.,  2010, \mn@doi [\mnras]
  {10.1111/j.1365-2966.2009.16087.x}, \href
  {http://adsabs.harvard.edu/abs/2010MNRAS.403...17J} {403, 17}

\bibitem[\protect\citeauthoryear{{Katz} \& {Gunn}}{{Katz} \&
  {Gunn}}{1991}]{1991ApJ...377..365K}
{Katz} N.,  {Gunn} J.~E.,  1991, \mn@doi [\apj] {10.1086/170367}, \href
  {http://adsabs.harvard.edu/abs/1991ApJ...377..365K} {377, 365}

\bibitem[\protect\citeauthoryear{{Katz}, {Weinberg}  \& {Hernquist}}{{Katz}
  et~al.}{1996}]{1996ApJS..105...19K}
{Katz} N.,  {Weinberg} D.~H.,   {Hernquist} L.,  1996, \mn@doi [\apjs]
  {10.1086/192305}, \href {http://adsabs.harvard.edu/abs/1996ApJS..105...19K}
  {105, 19}

\bibitem[\protect\citeauthoryear{{Kormendy} \& {Ho}}{{Kormendy} \&
  {Ho}}{2013}]{2013ARA&A..51..511K}
{Kormendy} J.,  {Ho} L.~C.,  2013, \mn@doi [\araa]
  {10.1146/annurev-astro-082708-101811}, \href
  {http://adsabs.harvard.edu/abs/2013ARA%26A..51..511K} {51, 511}

\bibitem[\protect\citeauthoryear{{Kormendy} \& {Kennicutt}}{{Kormendy} \&
  {Kennicutt}}{2004}]{2004ARA&A..42..603K}
{Kormendy} J.,  {Kennicutt} Jr. R.~C.,  2004, \mn@doi [\araa]
  {10.1146/annurev.astro.42.053102.134024}, \href
  {http://adsabs.harvard.edu/abs/2004ARA%26A..42..603K} {42, 603}

\bibitem[\protect\citeauthoryear{{Leitherer} et~al.,}{{Leitherer}
  et~al.}{1999}]{1999ApJS..123....3L}
{Leitherer} C.,  et~al., 1999, \mn@doi [\apjs] {10.1086/313233}, \href
  {http://adsabs.harvard.edu/abs/1999ApJS..123....3L} {123, 3}

\bibitem[\protect\citeauthoryear{{Leitherer}, {Ortiz Ot{\'a}lvaro}, {Bresolin},
  {Kudritzki}, {Lo Faro}, {Pauldrach}, {Pettini}  \& {Rix}}{{Leitherer}
  et~al.}{2010}]{2010ApJS..189..309L}
{Leitherer} C.,  {Ortiz Ot{\'a}lvaro} P.~A.,  {Bresolin} F.,  {Kudritzki}
  R.-P.,  {Lo Faro} B.,  {Pauldrach} A.~W.~A.,  {Pettini} M.,   {Rix} S.~A.,
  2010, \mn@doi [\apjs] {10.1088/0067-0049/189/2/309}, \href
  {http://adsabs.harvard.edu/abs/2010ApJS..189..309L} {189, 309}

\bibitem[\protect\citeauthoryear{{MacArthur}, {Courteau}  \&
  {Holtzman}}{{MacArthur} et~al.}{2003}]{2003ApJ...582..689M}
{MacArthur} L.~A.,  {Courteau} S.,   {Holtzman} J.~A.,  2003, \mn@doi [\apj]
  {10.1086/344506}, \href {http://adsabs.harvard.edu/abs/2003ApJ...582..689M}
  {582, 689}

\bibitem[\protect\citeauthoryear{{MacArthur}, {Ellis}, {Treu}, {U}, {Bundy}  \&
  {Moran}}{{MacArthur} et~al.}{2008}]{2008ApJ...680...70M}
{MacArthur} L.~A.,  {Ellis} R.~S.,  {Treu} T.,  {U} V.,  {Bundy} K.,   {Moran}
  S.,  2008, \mn@doi [\apj] {10.1086/587887}, \href
  {http://adsabs.harvard.edu/abs/2008ApJ...680...70M} {680, 70}

\bibitem[\protect\citeauthoryear{{Marinacci}, {Pakmor}  \&
  {Springel}}{{Marinacci} et~al.}{2014}]{2014MNRAS.437.1750M}
{Marinacci} F.,  {Pakmor} R.,   {Springel} V.,  2014, \mn@doi [\mnras]
  {10.1093/mnras/stt2003}, \href
  {http://adsabs.harvard.edu/abs/2014MNRAS.437.1750M} {437, 1750}

\bibitem[\protect\citeauthoryear{{McCarthy}, {Font}, {Crain}, {Deason},
  {Schaye}  \& {Theuns}}{{McCarthy} et~al.}{2012a}]{2012MNRAS.420.2245M}
{McCarthy} I.~G.,  {Font} A.~S.,  {Crain} R.~A.,  {Deason} A.~J.,  {Schaye} J.,
    {Theuns} T.,  2012a, \mn@doi [\mnras] {10.1111/j.1365-2966.2011.20189.x},
  \href {http://adsabs.harvard.edu/abs/2012MNRAS.420.2245M} {420, 2245}

\bibitem[\protect\citeauthoryear{{McCarthy}, {Schaye}, {Font}, {Theuns},
  {Frenk}, {Crain}  \& {Dalla Vecchia}}{{McCarthy}
  et~al.}{2012b}]{2012MNRAS.427..379M}
{McCarthy} I.~G.,  {Schaye} J.,  {Font} A.~S.,  {Theuns} T.,  {Frenk} C.~S.,
  {Crain} R.~A.,   {Dalla Vecchia} C.,  2012b, \mn@doi [\mnras]
  {10.1111/j.1365-2966.2012.21951.x}, \href
  {http://adsabs.harvard.edu/abs/2012MNRAS.427..379M} {427, 379}

\bibitem[\protect\citeauthoryear{{McKinnon}, {Torrey}  \&
  {Vogelsberger}}{{McKinnon} et~al.}{2016}]{2016MNRAS.457.3775M}
{McKinnon} R.,  {Torrey} P.,   {Vogelsberger} M.,  2016, \mn@doi [\mnras]
  {10.1093/mnras/stw253}, \href
  {http://adsabs.harvard.edu/abs/2016MNRAS.457.3775M} {457, 3775}

\bibitem[\protect\citeauthoryear{{McQuinn}, {Lidz}, {Zaldarriaga}, {Hernquist},
  {Hopkins}, {Dutta}  \& {Faucher-Gigu{\`e}re}}{{McQuinn}
  et~al.}{2009}]{2009ApJ...694..842M}
{McQuinn} M.,  {Lidz} A.,  {Zaldarriaga} M.,  {Hernquist} L.,  {Hopkins} P.~F.,
   {Dutta} S.,   {Faucher-Gigu{\`e}re} C.-A.,  2009, \mn@doi [\apj]
  {10.1088/0004-637X/694/2/842}, \href
  {http://adsabs.harvard.edu/abs/2009ApJ...694..842M} {694, 842}

\bibitem[\protect\citeauthoryear{{Mendel}, {Simard}, {Palmer}, {Ellison}  \&
  {Patton}}{{Mendel} et~al.}{2014}]{2014ApJS..210....3M}
{Mendel} J.~T.,  {Simard} L.,  {Palmer} M.,  {Ellison} S.~L.,   {Patton} D.~R.,
   2014, \mn@doi [\apjs] {10.1088/0067-0049/210/1/3}, \href
  {http://adsabs.harvard.edu/abs/2014ApJS..210....3M} {210, 3}

\bibitem[\protect\citeauthoryear{{Micha{\l}owski}, {Hayward}, {Dunlop},
  {Bruce}, {Cirasuolo}, {Cullen}  \& {Hernquist}}{{Micha{\l}owski}
  et~al.}{2014}]{2014A&A...571A..75M}
{Micha{\l}owski} M.~J.,  {Hayward} C.~C.,  {Dunlop} J.~S.,  {Bruce} V.~A.,
  {Cirasuolo} M.,  {Cullen} F.,   {Hernquist} L.,  2014, \mn@doi [\aap]
  {10.1051/0004-6361/201424174}, \href
  {http://adsabs.harvard.edu/abs/2014A%26A...571A..75M} {571, A75}

\bibitem[\protect\citeauthoryear{{Miller}, {Bundy}, {Sullivan}, {Ellis}  \&
  {Treu}}{{Miller} et~al.}{2011}]{2011ApJ...741..115M}
{Miller} S.~H.,  {Bundy} K.,  {Sullivan} M.,  {Ellis} R.~S.,   {Treu} T.,
  2011, \mn@doi [\apj] {10.1088/0004-637X/741/2/115}, \href
  {http://adsabs.harvard.edu/abs/2011ApJ...741..115M} {741, 115}

\bibitem[\protect\citeauthoryear{{Mitchell}, {Lacey}, {Baugh}  \&
  {Cole}}{{Mitchell} et~al.}{2013}]{2013MNRAS.435...87M}
{Mitchell} P.~D.,  {Lacey} C.~G.,  {Baugh} C.~M.,   {Cole} S.,  2013, \mn@doi
  [\mnras] {10.1093/mnras/stt1280}, \href
  {http://adsabs.harvard.edu/abs/2013MNRAS.435...87M} {435, 87}

\bibitem[\protect\citeauthoryear{{Moustakas} et~al.,}{{Moustakas}
  et~al.}{2013}]{2013ApJ...767...50M}
{Moustakas} J.,  et~al., 2013, \mn@doi [\apj] {10.1088/0004-637X/767/1/50},
  \href {http://adsabs.harvard.edu/abs/2013ApJ...767...50M} {767, 50}

\bibitem[\protect\citeauthoryear{{Munshi} et~al.,}{{Munshi}
  et~al.}{2013}]{2013ApJ...766...56M}
{Munshi} F.,  et~al., 2013, \mn@doi [\apj] {10.1088/0004-637X/766/1/56}, \href
  {http://adsabs.harvard.edu/abs/2013ApJ...766...56M} {766, 56}

\bibitem[\protect\citeauthoryear{{Nair} \& {Abraham}}{{Nair} \&
  {Abraham}}{2010}]{2010ApJS..186..427N}
{Nair} P.~B.,  {Abraham} R.~G.,  2010, \mn@doi [\apjs]
  {10.1088/0067-0049/186/2/427}, \href
  {http://adsabs.harvard.edu/abs/2010ApJS..186..427N} {186, 427}

\bibitem[\protect\citeauthoryear{{Navarro} \& {Steinmetz}}{{Navarro} \&
  {Steinmetz}}{1997}]{1997ApJ...478...13N}
{Navarro} J.~F.,  {Steinmetz} M.,  1997, \apj, \href
  {http://adsabs.harvard.edu/abs/1997ApJ...478...13N} {478, 13}

\bibitem[\protect\citeauthoryear{{Navarro} \& {Steinmetz}}{{Navarro} \&
  {Steinmetz}}{2000}]{2000ApJ...538..477N}
{Navarro} J.~F.,  {Steinmetz} M.,  2000, \mn@doi [\apj] {10.1086/309175}, \href
  {http://adsabs.harvard.edu/abs/2000ApJ...538..477N} {538, 477}

\bibitem[\protect\citeauthoryear{{Navarro} \& {White}}{{Navarro} \&
  {White}}{1994}]{1994MNRAS.267..401N}
{Navarro} J.~F.,  {White} S.~D.~M.,  1994, \mn@doi [\mnras]
  {10.1093/mnras/267.2.401}, \href
  {http://adsabs.harvard.edu/abs/1994MNRAS.267..401N} {267, 401}

\bibitem[\protect\citeauthoryear{{Navarro}, {Frenk}  \& {White}}{{Navarro}
  et~al.}{1995}]{1995MNRAS.275...56N}
{Navarro} J.~F.,  {Frenk} C.~S.,   {White} S.~D.~M.,  1995, \mn@doi [\mnras]
  {10.1093/mnras/275.1.56}, \href
  {http://adsabs.harvard.edu/abs/1995MNRAS.275...56N} {275, 56}

\bibitem[\protect\citeauthoryear{{Nelson} et~al.,}{{Nelson}
  et~al.}{2015}]{2015A&C....13...12N}
{Nelson} D.,  et~al., 2015, \mn@doi [Astronomy and Computing]
  {10.1016/j.ascom.2015.09.003}, \href
  {http://adsabs.harvard.edu/abs/2015A%26C....13...12N} {13, 12}

\bibitem[\protect\citeauthoryear{{Obreja}, {Stinson}, {Dutton}, {Macci{\`o}},
  {Wang}  \& {Kang}}{{Obreja} et~al.}{2016}]{2016MNRAS.459..467O}
{Obreja} A.,  {Stinson} G.~S.,  {Dutton} A.~A.,  {Macci{\`o}} A.~V.,  {Wang}
  L.,   {Kang} X.,  2016, \mn@doi [\mnras] {10.1093/mnras/stw690}, \href
  {http://adsabs.harvard.edu/abs/2016MNRAS.459..467O} {459, 467}

\bibitem[\protect\citeauthoryear{{Oesch} et~al.,}{{Oesch}
  et~al.}{2013}]{2013ApJ...773...75O}
{Oesch} P.~A.,  et~al., 2013, \mn@doi [\apj] {10.1088/0004-637X/773/1/75},
  \href {http://adsabs.harvard.edu/abs/2013ApJ...773...75O} {773, 75}

\bibitem[\protect\citeauthoryear{{Okamoto}, {Eke}, {Frenk}  \&
  {Jenkins}}{{Okamoto} et~al.}{2005}]{2005MNRAS.363.1299O}
{Okamoto} T.,  {Eke} V.~R.,  {Frenk} C.~S.,   {Jenkins} A.,  2005, \mn@doi
  [\mnras] {10.1111/j.1365-2966.2005.09525.x}, \href
  {http://adsabs.harvard.edu/abs/2005MNRAS.363.1299O} {363, 1299}

\bibitem[\protect\citeauthoryear{Piontek \& Steinmetz}{Piontek \&
  Steinmetz}{2009}]{piontek2009angular}
Piontek F.,  Steinmetz M.,  2009, arXiv preprint arXiv:0909.4156

\bibitem[\protect\citeauthoryear{{Sales}, {Navarro}, {Schaye}, {Dalla Vecchia},
  {Springel}  \& {Booth}}{{Sales} et~al.}{2010}]{2010MNRAS.409.1541S}
{Sales} L.~V.,  {Navarro} J.~F.,  {Schaye} J.,  {Dalla Vecchia} C.,  {Springel}
  V.,   {Booth} C.~M.,  2010, \mn@doi [\mnras]
  {10.1111/j.1365-2966.2010.17391.x}, \href
  {http://adsabs.harvard.edu/abs/2010MNRAS.409.1541S} {409, 1541}

\bibitem[\protect\citeauthoryear{{Scannapieco}, {White}, {Springel}  \&
  {Tissera}}{{Scannapieco} et~al.}{2009}]{2009MNRAS.396..696S}
{Scannapieco} C.,  {White} S.~D.~M.,  {Springel} V.,   {Tissera} P.~B.,  2009,
  \mn@doi [\mnras] {10.1111/j.1365-2966.2009.14764.x}, \href
  {http://adsabs.harvard.edu/abs/2009MNRAS.396..696S} {396, 696}

\bibitem[\protect\citeauthoryear{{Scannapieco}, {Gadotti}, {Jonsson}  \&
  {White}}{{Scannapieco} et~al.}{2010}]{2010MNRAS.407L..41S}
{Scannapieco} C.,  {Gadotti} D.~A.,  {Jonsson} P.,   {White} S.~D.~M.,  2010,
  \mn@doi [\mnras] {10.1111/j.1745-3933.2010.00900.x}, \href
  {http://adsabs.harvard.edu/abs/2010MNRAS.407L..41S} {407, L41}

\bibitem[\protect\citeauthoryear{{Scannapieco} et~al.,}{{Scannapieco}
  et~al.}{2012}]{2012MNRAS.423.1726S}
{Scannapieco} C.,  et~al., 2012, \mn@doi [\mnras]
  {10.1111/j.1365-2966.2012.20993.x}, \href
  {http://adsabs.harvard.edu/abs/2012MNRAS.423.1726S} {423, 1726}

\bibitem[\protect\citeauthoryear{{Schade}, {Carlberg}, {Yee}, {Lopez-Cruz}  \&
  {Ellingson}}{{Schade} et~al.}{1996}]{1996ApJ...464L..63S}
{Schade} D.,  {Carlberg} R.~G.,  {Yee} H.~K.~C.,  {Lopez-Cruz} O.,
  {Ellingson} E.,  1996, \mn@doi [\apjl] {10.1086/310091}, \href
  {http://adsabs.harvard.edu/abs/1996ApJ...464L..63S} {464, L63}

\bibitem[\protect\citeauthoryear{{Schaye} et~al.,}{{Schaye}
  et~al.}{2015}]{2015MNRAS.446..521S}
{Schaye} J.,  et~al., 2015, \mn@doi [\mnras] {10.1093/mnras/stu2058}, \href
  {http://adsabs.harvard.edu/abs/2015MNRAS.446..521S} {446, 521}

\bibitem[\protect\citeauthoryear{{Shen}, {Mo}, {White}, {Blanton}, {Kauffmann},
  {Voges}, {Brinkmann}  \& {Csabai}}{{Shen} et~al.}{2003}]{2003MNRAS.343..978S}
{Shen} S.,  {Mo} H.~J.,  {White} S.~D.~M.,  {Blanton} M.~R.,  {Kauffmann} G.,
  {Voges} W.,  {Brinkmann} J.,   {Csabai} I.,  2003, \mn@doi [\mnras]
  {10.1046/j.1365-8711.2003.06740.x}, \href
  {http://adsabs.harvard.edu/abs/2003MNRAS.343..978S} {343, 978}

\bibitem[\protect\citeauthoryear{{Sijacki} \& {Springel}}{{Sijacki} \&
  {Springel}}{2006}]{2006MNRAS.366..397S}
{Sijacki} D.,  {Springel} V.,  2006, \mn@doi [\mnras]
  {10.1111/j.1365-2966.2005.09860.x}, \href
  {http://adsabs.harvard.edu/abs/2006MNRAS.366..397S} {366, 397}

\bibitem[\protect\citeauthoryear{{Sijacki}, {Springel}, {Di Matteo}  \&
  {Hernquist}}{{Sijacki} et~al.}{2007}]{2007MNRAS.380..877S}
{Sijacki} D.,  {Springel} V.,  {Di Matteo} T.,   {Hernquist} L.,  2007, \mn@doi
  [\mnras] {10.1111/j.1365-2966.2007.12153.x}, \href
  {http://adsabs.harvard.edu/abs/2007MNRAS.380..877S} {380, 877}

\bibitem[\protect\citeauthoryear{{Simard}}{{Simard}}{1998}]{1998ASPC..145..108S}
{Simard} L.,  1998, in {Albrecht} R.,  {Hook} R.~N.,   {Bushouse} H.~A.,  eds,
  Astronomical Society of the Pacific Conference Series Vol. 145, Astronomical
  Data Analysis Software and Systems VII. p.~108

\bibitem[\protect\citeauthoryear{{Simard} et~al.,}{{Simard}
  et~al.}{2002}]{2002ApJS..142....1S}
{Simard} L.,  et~al., 2002, \mn@doi [\apjs] {10.1086/341399}, \href
  {http://adsabs.harvard.edu/abs/2002ApJS..142....1S} {142, 1}

\bibitem[\protect\citeauthoryear{{Simard}, {Mendel}, {Patton}, {Ellison}  \&
  {McConnachie}}{{Simard} et~al.}{2011}]{2011ApJS..196...11S}
{Simard} L.,  {Mendel} J.~T.,  {Patton} D.~R.,  {Ellison} S.~L.,
  {McConnachie} A.~W.,  2011, \mn@doi [\apjs] {10.1088/0067-0049/196/1/11},
  \href {http://adsabs.harvard.edu/abs/2011ApJS..196...11S} {196, 11}

\bibitem[\protect\citeauthoryear{{Snyder} et~al.,}{{Snyder}
  et~al.}{2015}]{2015MNRAS.454.1886S}
{Snyder} G.~F.,  et~al., 2015, \mn@doi [\mnras] {10.1093/mnras/stv2078}, \href
  {http://adsabs.harvard.edu/abs/2015MNRAS.454.1886S} {454, 1886}

\bibitem[\protect\citeauthoryear{{Sparre} \& {Springel}}{{Sparre} \&
  {Springel}}{2016}]{2016arXiv160408205S}
{Sparre} M.,  {Springel} V.,  2016, preprint, \href
  {http://adsabs.harvard.edu/abs/2016arXiv160408205S} {} (\mn@eprint {arXiv}
  {1604.08205})

\bibitem[\protect\citeauthoryear{{Springel}}{{Springel}}{2010}]{2010MNRAS.401..791S}
{Springel} V.,  2010, \mn@doi [\mnras] {10.1111/j.1365-2966.2009.15715.x},
  \href {http://adsabs.harvard.edu/abs/2010MNRAS.401..791S} {401, 791}

\bibitem[\protect\citeauthoryear{{Springel} \& {Hernquist}}{{Springel} \&
  {Hernquist}}{2003}]{2003MNRAS.339..289S}
{Springel} V.,  {Hernquist} L.,  2003, \mn@doi [\mnras]
  {10.1046/j.1365-8711.2003.06206.x}, \href
  {http://adsabs.harvard.edu/abs/2003MNRAS.339..289S} {339, 289}

\bibitem[\protect\citeauthoryear{{Springel}, {White}  \&
  {Hernquist}}{{Springel} et~al.}{2001}]{2001ApJ...549..681S}
{Springel} V.,  {White} M.,   {Hernquist} L.,  2001, \mn@doi [\apj]
  {10.1086/319473}, \href {http://adsabs.harvard.edu/abs/2001ApJ...549..681S}
  {549, 681}

\bibitem[\protect\citeauthoryear{{Springel}, {Di Matteo}  \&
  {Hernquist}}{{Springel} et~al.}{2005}]{2005MNRAS.361..776S}
{Springel} V.,  {Di Matteo} T.,   {Hernquist} L.,  2005, \mn@doi [\mnras]
  {10.1111/j.1365-2966.2005.09238.x}, \href
  {http://adsabs.harvard.edu/abs/2005MNRAS.361..776S} {361, 776}

\bibitem[\protect\citeauthoryear{{Steinmetz} \& {Navarro}}{{Steinmetz} \&
  {Navarro}}{1999}]{1999ApJ...513..555S}
{Steinmetz} M.,  {Navarro} J.~F.,  1999, \mn@doi [\apj] {10.1086/306904}, \href
  {http://adsabs.harvard.edu/abs/1999ApJ...513..555S} {513, 555}

\bibitem[\protect\citeauthoryear{{Stinson}, {Bailin}, {Couchman}, {Wadsley},
  {Shen}, {Nickerson}, {Brook}  \& {Quinn}}{{Stinson}
  et~al.}{2010}]{2010MNRAS.408..812S}
{Stinson} G.~S.,  {Bailin} J.,  {Couchman} H.,  {Wadsley} J.,  {Shen} S.,
  {Nickerson} S.,  {Brook} C.,   {Quinn} T.,  2010, \mn@doi [\mnras]
  {10.1111/j.1365-2966.2010.17187.x}, \href
  {http://adsabs.harvard.edu/abs/2010MNRAS.408..812S} {408, 812}

\bibitem[\protect\citeauthoryear{{Torrey}, {Vogelsberger}, {Genel}, {Sijacki},
  {Springel}  \& {Hernquist}}{{Torrey} et~al.}{2014}]{2014MNRAS.438.1985T}
{Torrey} P.,  {Vogelsberger} M.,  {Genel} S.,  {Sijacki} D.,  {Springel} V.,
  {Hernquist} L.,  2014, \mn@doi [\mnras] {10.1093/mnras/stt2295}, \href
  {http://adsabs.harvard.edu/abs/2014MNRAS.438.1985T} {438, 1985}

\bibitem[\protect\citeauthoryear{{Torrey} et~al.,}{{Torrey}
  et~al.}{2015}]{2015MNRAS.447.2753T}
{Torrey} P.,  et~al., 2015, \mn@doi [\mnras] {10.1093/mnras/stu2592}, \href
  {http://adsabs.harvard.edu/abs/2015MNRAS.447.2753T} {447, 2753}

\bibitem[\protect\citeauthoryear{{Trayford} et~al.,}{{Trayford}
  et~al.}{2015}]{2015MNRAS.452.2879T}
{Trayford} J.~W.,  et~al., 2015, \mn@doi [\mnras] {10.1093/mnras/stv1461},
  \href {http://adsabs.harvard.edu/abs/2015MNRAS.452.2879T} {452, 2879}

\bibitem[\protect\citeauthoryear{{Trujillo} et~al.,}{{Trujillo}
  et~al.}{2004}]{2004ApJ...604..521T}
{Trujillo} I.,  et~al., 2004, \mn@doi [\apj] {10.1086/382060}, \href
  {http://adsabs.harvard.edu/abs/2004ApJ...604..521T} {604, 521}

\bibitem[\protect\citeauthoryear{{V{\'a}zquez} \& {Leitherer}}{{V{\'a}zquez} \&
  {Leitherer}}{2005}]{2005ApJ...621..695V}
{V{\'a}zquez} G.~A.,  {Leitherer} C.,  2005, \mn@doi [\apj] {10.1086/427866},
  \href {http://adsabs.harvard.edu/abs/2005ApJ...621..695V} {621, 695}

\bibitem[\protect\citeauthoryear{{Vogelsberger}, {Sijacki}, {Kere{\v s}},
  {Springel}  \& {Hernquist}}{{Vogelsberger}
  et~al.}{2012}]{2012MNRAS.425.3024V}
{Vogelsberger} M.,  {Sijacki} D.,  {Kere{\v s}} D.,  {Springel} V.,
  {Hernquist} L.,  2012, \mn@doi [\mnras] {10.1111/j.1365-2966.2012.21590.x},
  \href {http://adsabs.harvard.edu/abs/2012MNRAS.425.3024V} {425, 3024}

\bibitem[\protect\citeauthoryear{{Vogelsberger}, {Genel}, {Sijacki}, {Torrey},
  {Springel}  \& {Hernquist}}{{Vogelsberger}
  et~al.}{2013}]{2013MNRAS.436.3031V}
{Vogelsberger} M.,  {Genel} S.,  {Sijacki} D.,  {Torrey} P.,  {Springel} V.,
  {Hernquist} L.,  2013, \mn@doi [\mnras] {10.1093/mnras/stt1789}, \href
  {http://adsabs.harvard.edu/abs/2013MNRAS.436.3031V} {436, 3031}

\bibitem[\protect\citeauthoryear{{Vogelsberger} et~al.,}{{Vogelsberger}
  et~al.}{2014a}]{2014MNRAS.444.1518V}
{Vogelsberger} M.,  et~al., 2014a, \mn@doi [\mnras] {10.1093/mnras/stu1536},
  \href {http://adsabs.harvard.edu/abs/2014MNRAS.444.1518V} {444, 1518}

\bibitem[\protect\citeauthoryear{{Vogelsberger} et~al.,}{{Vogelsberger}
  et~al.}{2014b}]{2014Natur.509..177V}
{Vogelsberger} M.,  et~al., 2014b, \mn@doi [\nat] {10.1038/nature13316}, \href
  {http://adsabs.harvard.edu/abs/2014Natur.509..177V} {509, 177}

\bibitem[\protect\citeauthoryear{{Wiersma}, {Schaye}  \& {Smith}}{{Wiersma}
  et~al.}{2009a}]{2009MNRAS.393...99W}
{Wiersma} R.~P.~C.,  {Schaye} J.,   {Smith} B.~D.,  2009a, \mn@doi [\mnras]
  {10.1111/j.1365-2966.2008.14191.x}, \href
  {http://adsabs.harvard.edu/abs/2009MNRAS.393...99W} {393, 99}

\bibitem[\protect\citeauthoryear{{Wiersma}, {Schaye}, {Theuns}, {Dalla Vecchia}
   \& {Tornatore}}{{Wiersma} et~al.}{2009b}]{2009MNRAS.399..574W}
{Wiersma} R.~P.~C.,  {Schaye} J.,  {Theuns} T.,  {Dalla Vecchia} C.,
  {Tornatore} L.,  2009b, \mn@doi [\mnras] {10.1111/j.1365-2966.2009.15331.x},
  \href {http://adsabs.harvard.edu/abs/2009MNRAS.399..574W} {399, 574}

\bibitem[\protect\citeauthoryear{{van Zee}}{{van
  Zee}}{2000}]{2000AJ....119.2757V}
{van Zee} L.,  2000, \mn@doi [\aj] {10.1086/301378}, \href
  {http://adsabs.harvard.edu/abs/2000AJ....119.2757V} {119, 2757}

\makeatother
\end{thebibliography}





\bsp	
\label{lastpage}
\end{document}